\documentclass[article]{./macros/elsarticle_qh}

\usepackage[hmarginratio=1:1,top=32mm,left=20mm,columnsep=1cm]{geometry} %
\usepackage{bm}
\usepackage{amsmath}
\usepackage{relsize} 
\usepackage[svgnames]{xcolor} 
\usepackage{grffile} 
\usepackage{graphicx}

\usepackage{hyperref}
\hypersetup{
	colorlinks=true,                          
	linkcolor=DarkRed,
	citecolor=DarkRed,
	urlcolor=DarkRed  }

\usepackage{bm}
\usepackage{amsmath}
\usepackage{booktabs}
\usepackage{amssymb}
\usepackage{etoolbox}
\usepackage{mathtools}
\usepackage{gensymb}
\usepackage[section]{placeins} 
\usepackage{float}
\usepackage{subfig}
\usepackage[thinc]{esdiff}

\DeclareMathAlphabet{\mathsfbi}{OT1}{\sfdefault}{bx}{sl}
\DeclareMathVersion{sfletters}
\SetSymbolFont{letters}{sfletters}{OML}{ntxsfmi}{b}{it}

\makeatletter
\newcommand{\mathbfsbilow}[1]{%
	\text{\mathversion{sfletters}$\m@th#1$}%
}
\DeclareRobustCommand{\tensor}[1]{%
	\begingroup
	\ifcat\noexpand #1\relax
	\edef\greek@test{\detokenize{#1}}%
	\edef\greek@test{\expandafter\@cdr\greek@test\@nil}%
	\edef\greek@test{\expandafter\@car\greek@test\@nil}%
	\edef\x{\the\lccode\expandafter`\greek@test}%
	\edef\y{\number\expandafter`\greek@test}%
	\ifnum\x=\y\relax
	\mathbfsbilow{#1}%
	\else
	\mathsfbi{#1}%
	\fi
	\else
	\mathsfbi{#1}%
	\fi
	\endgroup
}
\makeatother
\makeatletter
\newcommand{\sbullet}{%
	\hbox{\fontfamily{lmr}\fontsize{.4\dimexpr(\f@size pt)}{0}\selectfont\textbullet}}

\makeatother

\usepackage[numbers]{natbib}

\begin{document}
	
\begin{frontmatter}
		
\title{{\Large \textbf{Nearly constant \textit{Q} dissipative models and wave equations for general viscoelastic anisotropy}}
}

\cortext[mycorrespondingauthor]{Corresponding author}

\address[KFUPM]{CPG, KFUPM, Dhahran, 31261, Saudi Arabia}
\address[ETHZ]{Institute of Geophysics, ETH Zurich, Zurich, 8092, Switzerland}

\author[KFUPM]{Qi Hao\corref{mycorrespondingauthor}} 
\ead{{xqi.hao@gmail.com, qi.hao@kfupm.edu.sa}}

\author[ETHZ]{Stewart Greenhalgh}
\ead{gstewart@retired.ethz.ch}

\begin{abstract}
The quality factor ($Q$) links seismic wave energy dissipation to physical properties of the Earth's interior, such as temperature, stress and composition. Frequency independence of $Q$, also called constant $Q$ for brevity, is a common assumption in practice for seismic $Q$ inversions. Although exactly and nearly constant $Q$ dissipative models are proposed in the literature, it is inconvenient to obtain constant $Q$ wave equations in differential form, which explicitly involve a specified $Q$ parameter. In our recent research paper, we proposed a novel weighting function method to build the first- and second-order nearly constant $Q$ dissipative models. Of importance is the fact that the wave equations in differential form  for these two models explicitly involve a specified $Q$ parameter. This behavior is beneficial for time-domain seismic waveform inversion for $Q$, which requires the first derivative of wavefields with respect to $Q$ parameters. In this paper, we extend the first- and second-order nearly constant $Q$ models to the general viscoelastic anisotropic case. We also present a few formulation of the nearly constant $Q$ viscoelastic anisotropic wave equations in differential form.  
\end{abstract}


\end{frontmatter}


\section{Introduction}
Seismic wave propagation in the Earth's interior is dissipative due to intrinsic anelasticity and small-scale heterogeneities. Observations of seismic wave attenuation provide information about temperature, stress, composition, fluid content and the defect nature of the solid Earth \cite{anderson:2007}. The quality factor, $Q$, describes the anelastic behavior of a dissipative medium. Although many seismological observations demonstrate the frequency dependence of $Q$, constant $Q$ (also called frequency independence of $Q$), is still a common assumption to develop an inverse method for $Q$ in exploration and global seismology \cite{sheriff:1995,shearer:2019}. 

Constant $Q$ can be analytically treated by two well known dissipative models. The Kolsky model \cite{kolsky:1956} and the Kjartansson model \cite{kjartansson:1979} are nearly constant $Q$ and exactly constant $Q$, respectively, under the definition of the quality factor suggested by \cite{connell:1978}. However, it is difficult to apply these two models to time-domain seismic wavefield forward and inverse modeling because of the temporal convolution-type constitutive relationship between stress and strain. The temporal convolution implies that the complete time history of the wavefield is required to compute the wavefield at the next time, which is computationally costly.  

Constant $Q$ can be numerically modeled by using the generalized standard-linear-solid (SLS) model to fit a given $Q$ value over a specified frequency range of interest \cite{liu:1976,emmerich:1987,blanch:1995,blanc:2016}. The generalized SLS model (equivalent to the generalized Maxwell model) leads to the wave equation in differential form \cite{hao.greenhalgh:2019}, which can be solved by multiple time-domain numerical methods such as the finite difference method \cite{carcione:1988b}, the pseudospectral method \cite{carcione:1993}, the finite element method \cite{ham:2012} and the spectral element method \cite{komatitsch.trump:1999}. Almost all nearly constant $Q$ wave equations for the generalized SLS model implicitly involve a specified $Q$ parameter \cite{emmerich:1987,blanch:1995,bohlen:2002}, because fitting a given $Q$ value requires numerically solving a nonlinear inverse problem for the relaxation times. As an exception, \cite{fichtner:2014} improved the $\tau$-method \cite{blanch:1995} to determine the generalized SLS model for nearly constant $Q$ and power-law $Q$. Their method still needs to fit a given $Q$ value or function in a frequency range of interest, but the wave equations from their method involve an explicit $Q$ parameter. 

Unlike all the methods mentioned above, we recently proposed a $Q$-independent weighting function method to build two nearly constant $Q$ models of the generalized SLS type in the viscoacoustic case \cite{hao.greenhalgh:2021}. The complex moduli for these two models are of the first- and second-order with respect to the inverse of a specified $Q$ parameter and hence they are called the first- and second-order nearly constant $Q$ models, respectively. The resulting viscoacoustic wave equations involve explicitly a specified $Q$ parameter, which is beneficial for waveform inversion. 

The aim of this paper is to extend the first- and second-order nearly constant $Q$ models \cite{hao.greenhalgh:2021} to general viscoelastic anisotropy and derive the corresponding wave equations in differential form. We start with the general viscoelastic anisotropic constitutive relations, which describe the relationship between stress and strain in the time and frequency domains. We then show the complex stiffness coefficients, relaxation functions and creep functions for the first- and second-order nearly constant $Q$ models in the general viscoelastic anisotropic case. We also provide a numerical example to demonstrate the nearly constant $Q$ behavior of these two dissipative models. From the wave equations for a general viscoelastic anisotropic medium, we use the newly proposed dissipative models to derive the corresponding wave equations in differential form, which explicitly involve the specified $Q$ parameters. 

Because we will frequently switch between the time- and frequency-domains, the definitions for the Fourier transform and its inverse are shown below for clarity. 

The Fourier transform of a temporal signal $f(t)$ is written as
\begin{equation} \label{eq:Fourier}
\hat{f}(\omega) = 
                  \int_{-\infty}^{\infty} 
                  f(t) e^{i \omega t}
                  \text{d}t ,
\end{equation}
where $t$ is time and $\omega$ is angular frequency.

The inverse Fourier transform of the frequency-domain signal $\hat{f}(\omega)$ is written as
\begin{equation} \label{eq:invFourier}
f(t) = \frac{1}{2\pi} 
       \int_{-\infty}^{\infty} 
       \hat{f}(\omega) 
       e^{- i \omega t}
       \text{d}\omega .
\end{equation}

As a consequence of the above definitions, the first temporal derivative ``$d/dt$'' corresponds to ``$-i\omega$'' in 
the frequency domain. 

\section{Constitutive relations for general viscoelastic anisotropy}
In this section, we show the constitutive relations in both the time and frequency domain for a general viscoelastic anisotropic medium. 

\subsection{Time-domain relations}
In a general viscoelastic anisotropic medium, the time-domain constitutive relationship between stress $\mathbf{s}(t)$ and strain $\mathbf{e}(t)$ is expressed by the Riemann-Stieltjes convolution integral \cite{gurtin:1962,apostol:1974}, namely
\begin{equation} \label{eq:const_t1}
\mathbf{s}(t) = \bm{\Psi}(t) \odot \mathbf{e}(t) ,
\end{equation}
where $\mathbf{s} = (\sigma_{xx}, \sigma_{yy}, \sigma_{zz}, \sigma_{yz}, \sigma_{xz}, \sigma_{xy})^T$ and $\mathbf{e} = (\epsilon_{xx}, \epsilon_{yy}, \epsilon_{zz}, 2\epsilon_{yz}, 2\epsilon_{xz}, 2\epsilon_{xy})^T$, respectively. Quantities $\sigma_{ij}$ and $\epsilon_{ij}$ are the components of the stress and strain tensors, respectively. Quantity $\bm{\Psi}$ denotes the relaxation function matrix, which is a 6 by 6 symmetric matrix. The operator ``$\odot$'' is defined as
\begin{equation} \label{eq:constit}
\bm{\Psi}(t) \odot \mathbf{e}(t)  = 
\int_{-\infty}^{t} \bm{\Psi}(t-\tau) \text{d} \mathbf{e}(\tau) .
\end{equation}
This equation already implies that the relaxation function is causal, viz., zero for negative time.

The relaxation function matrix physically means the stress response corresponding to a unit step function (the Heaviside step function) in strain, starting at zero time, that is taking account of $\mathbf{e} = \mathbf{I} H(t)$, where $\mathbf{I}$ and $H(t)$ denote the identity matrix and the Heaviside step function, respectively. If the viscoelastic anisotropic medium is designated to start moving at $t=0$, the stress and strain in equation \ref{eq:const_t1} are non-zero for a positive time ($t>0$) and zero for a negative time ($t<0$). Hence, the constitutive equation \ref{eq:const_t1} can be rewritten as \cite{gurtin:1962,hudson:1980,hao.greenhalgh:2021}
\begin{equation} \label{eq:odot}
\bm{\Psi}(t) \odot \mathbf{e}(t) = 
\breve{\bm{\Psi}}(0+) \bm{e}(t) + \dot{\bm{\Psi}}(t) * \mathbf{e}(t), 
\end{equation}
where ``$0+$'' means that time approaches zero from the positive axis. The dot on $\dot{\bm{\Psi}}$ denotes temporal derivative. Matrix $\breve{\bm{\Psi}}(0+)$ denotes the result after excluding the singularity in the elements of matrix $\bm{\Psi}(0+)$. Let $\breve{\Psi}_{IJ}(t)$ and $\Psi_{IJ}(t)$ denote the elements of matrices $\breve{\bm{\Psi}}(t)$ and $\bm{\Psi}(t)$, respectively, where $I,J=1,2,...,6$. In the case that $\Psi_{IJ}(0+)$ has no singularity, for example, for the first- and second-order nearly constant $Q$ models shown in the next section, then $\breve{\Psi}_{IJ}(0+) = \Psi_{IJ}(0+)$. In the case that $\Psi_{IJ}(0+)$ is singular, for example, for the Kjartansson model, then $\breve{\Psi}_{IJ}(0+) = 0$. 
The operator ``$*$'' denotes the temporal convolution defined as
\begin{equation} \label{eq:star}
\dot{\bm{\Psi}}(t) * \mathbf{e}(t) = 
\int_{0}^{t} \dot{\bm{\Psi}}(t-\tau) \mathbf{e}(\tau) \text{d}\tau . 
\end{equation}

As the inverse of equation \ref{eq:const_t1}, the constitutive relation for the strain as a function of the stress is written as
\begin{equation} \label{eq:const_t2}
\mathbf{e}(t) = \mathbf{X}(t) \odot \mathbf{s}(t) ,
\end{equation}
where $\mathbf{X}$ denotes the creep function matrix. It physically means the strain response corresponding to a unit step function in stress, starting at $t=0$. 

A combination of the physical meaning of the creep function and the constitutive relation \ref{eq:const_t1} leads to the relationship between the relaxation and creep function matrices
\begin{equation} \label{eq:PsiX}
\bm{\Psi}(t) \odot \mathbf{X}(t) = \mathbf{I} H(t) ,
\end{equation}
where $\mathbf{I}$ denotes the identity matrix and $H(t)$ denotes the Heaviside step function of time.

\subsection{Frequency-domain relations}
The Fourier transform of equation \ref{eq:const_t1} gives rise to the frequency-domain constitutive relation for the stress as a function of the strain
\begin{equation} \label{eq:const_f1}
\hat{\mathbf{s}}(\omega) = \mathbf{M}(\omega) \hat{\mathbf{e}}(\omega) ,
\end{equation}
where $\mathbf{M}(\omega)$ denotes the complex stiffness matrix given by
\begin{equation}  \label{eq:Momega}
\mathbf{M}(\omega) =  \breve{\bm{\Psi}}(0+) 
+ \int_{0}^{\infty} \dot{\bm{\Psi}}(t) e^{i\omega t} \text{d}t .
\end{equation}
Here, we have taken account of equations \ref{eq:odot} and \ref{eq:star}.

Transforming equation \ref{eq:const_t2} into the frequency domain, we obtain the constitutive relation for the strain as a function of the stress
\begin{equation} \label{eq:const_f2}
\hat{\mathbf{e}}(\omega) = \mathbf{J}(\omega) \hat{\mathbf{s}}(\omega) ,
\end{equation} 
where $\mathbf{J}$ denotes the complex compliance matrix. It physically means the strain response due to a sinusoidal stress of frequency $\omega$ and amplitude unity. 

By analogy with equation \ref{eq:Momega}, the complex compliance matrix is expressed in terms of the creep function matrix as
\begin{equation}  \label{eq:Jomega}
\mathbf{J}(\omega) =  \breve{\mathbf{X}}(0+) 
+ \int_{0}^{\infty} \dot{\mathbf{X}}(t) e^{i\omega t} \text{d}t .
\end{equation}

The relationship between the complex stiffness coefficient matrix and the compliance matrix is expressed as
\begin{equation} \label{eq:MJ}
\mathbf{M}(\omega) \mathbf{J}(\omega) = \mathbf{I} .
\end{equation} 

\section{Nearly constant $Q$ models for general viscoelastic anisotropy}
In this section, we extend the first- and second-order nearly constant $Q$ models \cite{hao.greenhalgh:2021} to the general viscoelastic anisotropic case. 

\subsection{Essential functions}
Referring to \cite{hao.greenhalgh:2021}, the $Q$-independent weighting function of the generalized SLS type is given by
\begin{equation} \label{eq:Weq}
W(\omega)
= \sum_{\ell=1}^{L} \frac{1-i\omega\tau_{\epsilon}^{(\ell)}}{1-i\omega\tau_{\sigma}^{(\ell)}} ,
\end{equation}
where $L$ denotes the total number of SLS elements. The minus sign in front of ``$i$'' corresponds to the sign convention in the exponential term of the Fourier transform (equation \ref{eq:Fourier}). Quantities $\tau_{\epsilon}^{(\ell)}$ and $\tau_{\sigma}^{(\ell)}$ are $Q$-independent strain and stress relaxation times in the $l$-th term (SLS element) in the summation for the weighting function, respectively. These relaxation times are determined by using the following equation
\begin{equation} \label{eq:W_Wr}
W(\omega) - W_{R}(\omega_{0})
= \sum_{\ell=1}^{L} \frac{1-i\omega\tau_{\epsilon}^{(\ell)}}{1-i\omega\tau_{\sigma}^{(\ell)}}
-
\sum_{\ell=1}^{L}
\frac{1 + \omega_{0}^2 \tau_{\epsilon}^{(\ell)} \tau_{\sigma}^{(\ell)}}
{1+\omega_{0}^2 \left(\tau_{\sigma}^{(\ell)}\right)^2} ,
\end{equation}
to fit a $Q$-independent term over a frequency range of interest from the complex stiffness coefficients for the Kolsky model and from the Maclaurin series expansion of those for the Kjartansson model (see Appendix A), namely
\begin{equation}
\frac{2}{\pi} \text{ln}\left| \frac{\omega}{\omega_{0}}\right| - i\text{sgn}(\omega) \approx W(\omega) - W_{R}(\omega_{0}) ,
\end{equation}
Here, $W_{R}(\omega)$ denotes the real part of $W(\omega)$. Quantity $\omega_{0}$ denotes reference angular frequency, which is set as the central frequency of the seismic source \cite{hao.greenhalgh:2021}. Such a fitting is transformed to an optimization problem, which depends only on a frequency range of interest. The optimal values of $\tau_{\epsilon}^{(\ell)}$ and $\tau_{\sigma}^{(\ell)}$ for various frequency ranges can be found in \cite{hao.greenhalgh:2021}. The complex stiffness coefficients for the Kolsky and Kjartansson models are shown in Appendix A.

From equations \ref{eq:Momega} and \ref{eq:Jomega}, we conclude that if function $W(\omega) - W_{R}(\omega_{0})$ is considered as a complex modulus/compliance then the corresponding relaxation/creep function $\zeta(t)$ is given by
\begin{equation} \label{eq:zeta}
\zeta(t) = - W_{R}(\omega_{0}) H(t) 
+ 
\sum_{\ell=1}^{L} 
\left[
1 - \left(1 - 
\frac{\tau_{\epsilon}^{(\ell)}}{\tau_{\sigma}^{(\ell)}}
\right)
e^{-\frac{t}{\tau_{\sigma}^{(\ell)}}}
\right] H(t) . 
\end{equation}

Function $\zeta(t)$ in equation \ref{eq:zeta} is rewritten as
\begin{equation} \label{eq:zeta_new}
\zeta(t) = g H(t) - \sum_{\ell=1}^{L}\xi^{(\ell)}(t) , 
\end{equation}
where $g$ is a constant given by
\begin{equation}
\label{eq:gnew}
g = \sum_{\ell=1}^{L} 
\frac{\frac{\tau_{\epsilon}^{(\ell)}}{\tau_{\sigma}^{(\ell)}}-1}
{1+\omega_{0}^2\left(\tau_{\sigma}^{(\ell)}\right)^2} , 
\end{equation}
and $\xi^{(\ell)}(t)$ is given by
\begin{equation}
\label{eq:xil}
\xi^{(\ell)}(t) = 
\left( 
\frac{\tau_{\epsilon}^{(\ell)}}{\tau_{\sigma}^{(\ell)}} - 1 
\right)
\left(
1 - e^{-\frac{t}{\tau_{\sigma}^{(\ell)}}}
\right) 
H(t). 
\end{equation}
Function $\xi^{(\ell)}(t)$ is zero at $t=0$ , namely $\xi^{(\ell)}(0)=0$. This leads to the following two properties
\begin{align}
\label{eq:xil_odot_eps}
& \xi^{(\ell)}(t) \odot \epsilon(t) = \dot{\xi}^{(\ell)}(t) * \epsilon(t) , \\ 
\label{eq:zeta_odot_eps}
& \zeta(t) \odot \epsilon(t) = g \epsilon(t) - \sum_{\ell=1}^{L} \dot{\xi}^{(\ell)}(t) * \epsilon(t) ,
\end{align}
where the dot on $\dot{\xi}(t)$ denotes the temporal derivative and the operation ``$*$'' denotes the temporal convolution defined in equation \ref{eq:star}.
Function $\dot{\xi}^{(\ell)}(t) * \epsilon(t)$ can be transformed to the following differential equation
\begin{equation} \label{eq:zetaxiprop}
\diffp{}{t} \left[\dot{\xi}^{(\ell)}(t)*\epsilon(t)\right] = s^{(\ell)} \epsilon(t) - \frac{1}{\tau_{\sigma}^{(\ell)}} \dot{\xi}^{(\ell)}(t)*\epsilon(t) ,
\end{equation}
where $s^{(\ell)}$ is given by
\begin{equation} \label{eq:sl}
s^{(\ell)} = \frac{1}{\tau_{\sigma}^{(\ell)}} 
\left( 
\frac{\tau_{\epsilon}^{(\ell)}}{\tau_{\sigma}^{(\ell)}} - 1 
\right) .
\end{equation}
The derivation of equation \ref{eq:zetaxiprop} can be found in \cite{hao.greenhalgh:2019}.

\subsection{The first- and second-order nearly constant $Q$ models}
As an extension of the viscoacoustic result of \cite{hao.greenhalgh:2021}, we show the first- and second-order nearly constant $Q$ models for general viscoelastic anisotropy below.

The complex stiffness matrix, relaxation function matrix and creep function matrix for the first-order nearly constant $Q$ model are expressed as
\begin{align} 
\label{eq:M1stgen}
& \mathbf{M}(\omega) = \mathbf{M}^{(0)} + \mathbf{M}^{(1)} \left[W(\omega) - W_{R}(\omega_{0}) \right] , \\
\label{eq:Psi1stgen}
& \bm{\Psi}(t) = \mathbf{M}^{(0)} H(t) + \mathbf{M}^{(1)} \zeta(t) , \\
\label{eq:X1stgen}
& \mathbf{X}(t) = \mathbf{J}^{(0)} \sum_{n=0}^{\infty} (-1)^n \left(\mathbf{K}^{(1)}\right)^n \zeta^{\langle n \rangle}(t) ,
\end{align}
The complex stiffness coefficient matrix, relaxation function matrix and creep function matrix for the second-order nearly constant $Q$ model are given by
\begin{align} 
\label{eq:M2ndgen}
& \mathbf{M}(\omega) = \mathbf{M}^{(0)} + \mathbf{M}^{(1)} \left[W(\omega) - W_{R}(\omega_{0}) \right] + \frac{1}{2} \mathbf{M}^{(2)} \left[W(\omega) - W_{R}(\omega_{0}) \right]^2, \\
\label{eq:Psi2ndgen}
& \bm{\Psi}(t) = \mathbf{M}^{(0)} H(t) + \mathbf{M}^{(1)} \zeta(t) + \frac{1}{2} \mathbf{M}^{(2)} \zeta^{\langle 2 \rangle}(t), \\
\label{eq:X2ndgen}
& \mathbf{X}(t) =  \mathbf{J}^{(0)} \sum_{n=0}^{\infty} (-1)^n
\left\{\mathbf{K}^{(1)} \zeta(t) 
+ \frac{1}{2} \mathbf{K}^{(2)} \zeta^{\langle 2 \rangle}(t) \right\}
^{\langle n \rangle} .
\end{align}

Superscript ``$\langle \cdot \rangle$'' in equations \ref{eq:X1stgen}, \ref{eq:Psi2ndgen} and \ref{eq:X2ndgen} is defined for a causal function $f(t)$ as
\begin{equation} \label{eq:fn}
f^{\langle n \rangle}(t) =
\begin{cases}
\underbrace{f(t) \odot f(t) \cdots \odot f(t)}_{n}, & \text{if } n > 1, \\
f(t), & \text{if } n = 1 , \\
H(t), & \text{if } n = 0 .
\end{cases}
\end{equation}

Creep function matrices \ref{eq:X1stgen} and \ref{eq:X2ndgen} are derived from the complex stiffness matrices \ref{eq:M1stgen} and \ref{eq:M2ndgen}, respectively, using equations \ref{eq:Jomega} and \ref{eq:MJ}, and the correspondence relationship between equations \ref{eq:W_Wr} and \ref{eq:zeta}. The derivation of these creep function matrices is given in Appendix B. Matrices $\mathbf{J}^{(0)}$, $\mathbf{K}^{(1)}$ and $\mathbf{K}^{(2)}$ in equations \ref{eq:X1stgen} and \ref{eq:X2ndgen} are given by
\begin{align}
& \mathbf{J}^{(0)} = \left(\mathbf{M}^{(0)} \right)^{-1} , \\
& \mathbf{K}^{(m)} = \mathbf{M}^{(m)} \mathbf{J}^{(0)} , \quad m=1,2. 
\end{align}

Matrix $\mathbf{M}^{(0)}$ denotes a matrix for the reference stiffness coefficients, the elements of which are denoted by $M_{IJ}^{(0)}$. 
Quantities $M_{IJ}^{(0)}$ are independent of the reference quality factors $Q_{IJ}$. Matrices $\mathbf{M}^{(1)}$ and $\mathbf{M}^{(2)}$ denote the first- and second-order matrices, respectively, with respect to the inverse of the quality factor parameters.  

Quantities $M_{IJ}^{(0)}$ and $Q_{IJ}$ parameterize the first- and second-order nearly constant $Q$ models. The non-zero independent elements in matrices $\mathbf{M}^{(0)}$, $\mathbf{M}^{(1)}$ and $\mathbf{M}^{(2)}$ are generally expressed as
\begin{equation}
\label{eq:MIJn}
M_{IJ}^{(n)} = \frac{M_{IJ}^{(0)}}{Q_{IJ}^{n}} , \quad n=0,1,2, 
\end{equation} 
where $M_{IJ}^{(n)}$ denote the elements of matrix $\mathbf{M}^{(n)}$. Quantities $Q_{IJ}^n$ at $n=0$ are designated as $Q_{IJ}^0 = 1$ even in the case of $Q_{IJ}=\infty$ (i.e. the non-dissipative case). For a specified symmetry type (class of anisotropy), matrix $M^{(n)}$ has the same pattern as the elastic stiffness matrix. Referring to \cite{musgrave:1970}, \cite{okaya:2003} and \cite{tsvankin:2012}, shown below are the explicit expressions for $\mathbf{M}^{(n)}$, $n=0,1,2$, in a few specific medium types.

In the \textit{viscoelastic isotropic case}, the expressions for $\mathbf{M}^{(n)}$ is given by
\begin{equation} \label{eq:Mniso}
\mathbf{M}^{(n)} =
\begingroup
\def\arraystretch{1.8}
\left(
\begin{matrix*}[c]
\frac{M_{P}^{(0)}}{Q_{P}^{n}}  & \frac{M_{P}^{(0)}}{Q_{P}^{n}}-2\frac{M_{S}^{(0)}}{Q_{S}^{n}} & \frac{M_{P}^{(0)}}{Q_{P}^{n}}-2\frac{M_{S}^{(0)}}{Q_{S}^{n}} & 0 & 0 & 0 \\
\frac{M_{P}^{(0)}}{Q_{P}^{n}}-2\frac{M_{S}^{(0)}}{Q_{S}^{n}} & \frac{M_{P}^{(0)}}{Q_{P}^{n}} & 
\frac{M_{P}^{(0)}}{Q_{P}^{n}}-2\frac{M_{S}^{(0)}}{Q_{S}^{n}} & 0 & 0 & 0 \\
\frac{M_{P}^{(0)}}{Q_{P}^{n}}-2\frac{M_{S}^{(0)}}{Q_{S}^{n}} & \frac{M_{P}^{(0)}}{Q_{P}^{n}}-2\frac{M_{S}^{(0)}}{Q_{S}^{n}} & 
\frac{M_{P}^{(0)}}{Q_{P}^{n}} & 0 & 0 & 0 \\
0        &     0     &        0        & \frac{M_{S}^{(0)}}{Q_{S}^{n}} & 0 & 0 \\
0        &     0     &        0        & 0 & \frac{M_{S}^{(0)}}{Q_{S}^{n}} & 0 \\
0        &     0     &        0        & 0 & 0 & \frac{M_{S}^{(0)}}{Q_{S}^{n}}
\end{matrix*}
\right)
\endgroup ,
\end{equation}
where $M_{P}^{(0)}$ and $M_{S}^{(0)}$ denote the reference bulk and shear moduli, respectively. Quantities $Q_{P}$ and $Q_{S}$ denote the reference quality factors of homogeneous plane P and S waves, respectively.

In the case of \textit{viscoelastic transverse isotropy with a vertical symmetry axis}, the expression for $\mathbf{M}^{(n)}$ is given by 
\begin{equation}
\mathbf{M}^{(n)} =
\begingroup
\def\arraystretch{1.8}
\left(
\begin{matrix*}[c]
\frac{M_{11}^{(0)}}{Q_{11}^{n}}  & \frac{M_{11}^{(0)}}{Q_{11}^{n}}-2\frac{M_{66}^{(0)}}{Q_{66}^{n}} & \frac{M_{13}^{(0)}}{Q_{13}^{n}} & 0 & 0 & 0 \\
\frac{M_{11}^{(0)}}{Q_{11}^{n}}-2\frac{M_{66}^{(0)}}{Q_{66}^{n}} & \frac{M_{11}^{(0)}}{Q_{11}^{n}} & 
\frac{M_{13}^{(0)}}{Q_{13}^{n}} & 0 & 0 & 0 \\
\frac{M_{13}^{(0)}}{Q_{13}^{n}} & \frac{M_{13}^{(0)}}{Q_{13}^{n}} & 
\frac{M_{33}^{(0)}}{Q_{33}^{n}} & 0 & 0 & 0 \\
0       &    0      &      0    & \frac{M_{55}^{(0)}}{Q_{55}^{n}} & 0 & 0 \\
0       &    0      &      0    & 0 & \frac{M_{55}^{(0)}}{Q_{55}^{n}} & 0 \\
0       &    0      &      0    & 0 & 0 & \frac{M_{66}^{(0)}}{Q_{66}^{n}}
\end{matrix*}
\right)
\endgroup .
\end{equation}

In the \textit{viscoelastic orthorhombic case}, the expression for $\mathbf{M}^{(n)}$ is given by
\begin{equation}
\mathbf{M}^{(n)} = 
\begingroup
\def\arraystretch{1.8}
\left(
\begin{matrix*}[c]
\frac{M_{11}^{(0)}}{Q_{11}^{n}} & \frac{M_{12}^{(0)}}{Q_{12}^{n}}  & \frac{M_{13}^{(0)}}{Q_{13}^{n}} & 0 & 0 & 0 \\
\frac{M_{12}^{(0)}}{Q_{12}^{n}} & \frac{M_{22}^{(0)}}{Q_{22}^{n}} & 
\frac{M_{23}^{(0)}}{Q_{23}^{n}} & 0 & 0 & 0 \\
\frac{M_{13}^{(0)}}{Q_{13}^{n}} & \frac{M_{23}^{(0)}}{Q_{23}^{n}} & 
\frac{M_{33}^{(0)}}{Q_{33}^{n}} & 0 & 0 & 0 \\
0         &    0      &    0    & \frac{M_{44}^{(0)}}{Q_{44}^{n}} & 0 & 0 \\
0         &    0      &    0    &  0  & \frac{M_{55}^{(0)}}{Q_{55}^{n}} & 0 \\
0         &    0      &    0    &  0  &  0 & \frac{M_{66}^{(0)}}{Q_{66}^{n}}
\end{matrix*}
\right) 
\endgroup . 
\end{equation}

In the \textit{viscoelastic monoclinic case}, the expression for $\mathbf{M}^{(n)}$ is given by
\begin{equation} \label{eq:Mnmono}
\mathbf{M}^{(n)} = 
\begingroup
\def\arraystretch{1.8}
\left(
\begin{matrix*}[c]
\frac{M_{11}^{(0)}}{Q_{11}^{n}} & \frac{M_{12}^{(0)}}{Q_{12}^{n}}  & \frac{M_{13}^{(0)}}{Q_{13}^{n}} & 0 & 0 & \frac{M_{16}^{(0)}}{Q_{16}^{n}} \\
\frac{M_{12}^{(0)}}{Q_{12}^{n}} & \frac{M_{22}^{(0)}}{Q_{22}^{n}} & 
\frac{M_{23}^{(0)}}{Q_{23}^{n}} & 0 & 0 & \frac{M_{26}^{(0)}}{Q_{16}^{n}} \\
\frac{M_{13}^{(0)}}{Q_{13}^{n}} & \frac{M_{23}^{(0)}}{Q_{23}^{n}} & 
\frac{M_{33}^{(0)}}{Q_{33}^{n}} & 0 & 0 & \frac{M_{36}^{(0)}}{Q_{16}^{n}} \\
0  &   0    &   0   & \frac{M_{44}^{(0)}}{Q_{44}^{n}} & \frac{M_{45}^{(0)}}{Q_{45}^{n}} & 0 \\
0  &   0    &   0   & \frac{M_{45}^{(0)}}{Q_{45}^{n}} & \frac{M_{55}^{(0)}}{Q_{55}^{n}} & 0 \\
\frac{M_{16}^{(0)}}{Q_{16}^{n}} &  \frac{M_{26}^{(0)}}{Q_{16}^{n}}    &  \frac{M_{36}^{(0)}}{Q_{16}^{n}} & 0 & 0 & \frac{M_{66}^{(0)}}{Q_{66}^{n}}
\end{matrix*}
\right) 
\endgroup , 
\end{equation}
where the symmetry plane of a monoclinic medium is orthogonal to the $z$ axis \cite{tsvankin:2012}.

The complex stiffness coefficients (i.e., elements of the complex stiffness matrix) for the first- and second-order nearly constant $Q$ models are approximations to those for the Kolsky and Kjartansson models, respectively. The second-order nearly constant $Q$ model can provide a more accurate approximation to exactly constant $Q$ than the first-order nearly constant $Q$ model, as demonstrated later in this section.

\subsection{Coordinate rotations}
We define a new coordinate system $(x',y',z')$ relative to the original coordinate system $(x, y, z)$. Both coordinate systems share the same coordinate origin. The relationship between these two coordinate systems is expressed as
\begin{equation}
\mathbf{x} = \bm{a} \mathbf{x}' ,
\end{equation}
where $\bm{a}$ denotes the coordinate rotation matrix given by
\begin{equation}
\bm{a} = 
\left(
\begin{matrix}
a_{11} & a_{12} & a_{13} \\ 
a_{21} & a_{22} & a_{23} \\           
a_{31} & a_{32} & a_{33}                  
\end{matrix}  
\right) .
\end{equation}
Here, the first, second and third columns of $\bm{a}$ describe the base vectors along the $x'$, $y'$ and $z'$ axes in the $(x, y, z)$ coordinate system, respectively. 

According to \cite{auld:1973}, the Bond transformation matrices are given by
\begin{align}
& \resizebox{0.9\linewidth}{!}{$
\mathbf{L} =
\left(
\begin{matrix}
a_{11}^2 & a_{12}^2 & a_{13}^2 & 2a_{12}a_{13} & 2a_{13}a_{11} & 2a_{11}a_{12} \\
a_{21}^2 & a_{22}^2 & a_{23}^2 & 2a_{22}a_{23} & 2a_{23}a_{21} & 2a_{21}a_{22} \\
a_{31}^2 & a_{32}^2 & a_{33}^2 & 2a_{32}a_{33} & 2a_{33}a_{31} & 2a_{31}a_{32} \\
a_{21}a_{31} & a_{22}a_{32} & a_{23}a_{33}  & a_{22}a_{33}+a_{23}a_{32} & a_{21}a_{33}+a_{23}a_{31} & a_{22}a_{31}+a_{21}a_{32} \\
a_{31}a_{11} & a_{32}a_{12} & a_{33}a_{13}  & a_{12}a_{33}+a_{13}a_{32} & a_{13}a_{31}+a_{11}a_{33} & a_{11}a_{32}+a_{12}a_{31} \\
a_{11}a_{21} & a_{12}a_{22} & a_{13}a_{23}  & a_{12}a_{23}+a_{13}a_{22} & a_{13}a_{21}+a_{11}a_{23} & a_{11}a_{22}+a_{12}a_{21}
\end{matrix}
\right) $}, \\
& \resizebox{0.9\linewidth}{!}{$
\mathbf{R} =
\left(
\begin{matrix}
a_{11}^2 & a_{12}^2 & a_{13}^2 & a_{12}a_{13} & a_{13}a_{11} & a_{11}a_{12} \\
a_{21}^2 & a_{22}^2 & a_{23}^2 & a_{22}a_{23} & a_{23}a_{21} & a_{21}a_{22} \\
a_{31}^2 & a_{32}^2 & a_{33}^2 & a_{32}a_{33} & a_{33}a_{31} & a_{31}a_{32} \\
2a_{21}a_{31} & 2a_{22}a_{32} & 2a_{23}a_{33}  & a_{22}a_{33}+a_{23}a_{32} & a_{21}a_{33}+a_{23}a_{31} & a_{22}a_{31}+a_{21}a_{32} \\
2a_{31}a_{11} & 2a_{32}a_{12} & 2a_{33}a_{13}  & a_{12}a_{33}+a_{13}a_{32} & a_{13}a_{31}+a_{11}a_{33} & a_{11}a_{32}+a_{12}a_{31} \\
2a_{11}a_{21} & 2a_{12}a_{22} & 2a_{13}a_{23}  & a_{12}a_{23}+a_{13}a_{22} & a_{13}a_{21}+a_{11}a_{23} & a_{11}a_{22}+a_{12}a_{21}
\end{matrix}
\right) $}.
\end{align}
Here, matrices $\mathbf{L}$ and $\mathbf{R}$ satisfy the following relationship
\begin{equation}
\mathbf{L}^T = \mathbf{R}^{-1} .
\end{equation}

For consistency with the previous section, the complex stiffness matrix, relaxation function matrix and creep function matrix in the original coordinate system $(x, y, z)$ are denoted by $\mathbf{M}(\omega)$, $\bm{\Psi}(t)$ and $\mathbf{X}(t)$, respectively. The complex stiffness matrix, relaxation function matrix and creep function matrix in the new coordinate system $(x', y', z')$ are defined as $\tilde{\mathbf{M}}(\omega)$, $\tilde{\bm{\Psi}}(t)$ and $\tilde{\mathbf{X}}(t)$, respectively. 

As known from equations \ref{eq:const_f1} and \ref{eq:const_f2}, the frequency-domain constitutive relations for a viscoelastic medium have the same form as those for an elastic medium. The complex stiffness matrix for a general viscoelastic anisotropic medium satisfies the same symmetry as the stiffness matrix for the corresponding elastic anisotropic medium. This relation is also true for the complex and real compliance matrices. Hence, the Bond transformation is applicable to the complex stiffness and compliance matrices. Following \cite{auld:1973}, we apply the Bond transformation to obtain the following equations
\begin{align}
\label{eq:M_rot}
& \mathbf{M}(\omega) = \mathbf{L} \tilde{\mathbf{M}}(\omega) \mathbf{L}^T , \\
\label{eq:J_rot}
& \mathbf{J}(\omega) = \mathbf{R} \tilde{\mathbf{J}}(\omega) \mathbf{R}^T . 
\end{align}
Likewise, the transformation relations for the relaxation and creep function matrices are given respectively by
\begin{align}
\label{eq:Psi_rot}
&\bm{\Psi}(t) = \mathbf{L} \tilde{\bm{\Psi}}(t) \mathbf{L}^T , \\
\label{eq:X_rot}
&\mathbf{X}(t) = \mathbf{R} \tilde{\mathbf{X}}(t) \mathbf{R}^T ,
\end{align} 

Equations \ref{eq:M_rot} through \ref{eq:X_rot} are valid for a general dissipative model. These rotations allow us to extend the treatment to arbitrary orientation of the axes of symmetry of the anisotropic medium. 

We next take account of the second-order nearly constant $Q$ model. By analogy with equations \ref{eq:M2ndgen} through \ref{eq:X2ndgen}, the complex stiffness matrix, relaxation function matrix and creep function matrix for the second-order nearly constant $Q$ model in the coordinate system $(x', y', z')$ are expressed as
\begin{align} 
\label{eq:M2ndgen_newsys}
& \tilde{\mathbf{M}}(\omega) = \tilde{\mathbf{M}}^{(0)} + \tilde{\mathbf{M}}^{(1)} \left[W(\omega) - W_{R}(\omega_{0}) \right] + \frac{1}{2} \tilde{\mathbf{M}}^{(2)} \left[W(\omega) - W_{R}(\omega_{0}) \right]^2 . \\
\label{eq:Psi2ndgen_newsys}
& \tilde{\bm{\Psi}}(t) = \tilde{\mathbf{M}}^{(0)} H(t) + \tilde{\mathbf{M}}^{(1)} \zeta(t) + \frac{1}{2} \tilde{\mathbf{M}}^{(2)} \zeta^{\langle 2 \rangle}(t),  \\
\label{eq:X2ndgen_newsys}
& \tilde{\mathbf{X}}(t) =  \tilde{\mathbf{J}}^{(0)} \sum_{n=0}^{\infty} (-1)^n
\left\{\tilde{\mathbf{K}}^{(1)} \zeta(t) 
+ \frac{1}{2} \tilde{\mathbf{K}}^{(2)} \zeta^{\langle 2 \rangle}(t) \right\}
^{\langle n \rangle} ,
\end{align}
with
\begin{align}
& \tilde{\mathbf{J}}^{(0)} = \left(\tilde{\mathbf{M}}^{(0)} \right)^{-1} , \\
& \tilde{\mathbf{K}}^{(m)} = \tilde{\mathbf{M}}^{(m)} \tilde{\mathbf{J}}^{(0)} , \quad m=1,2. 
\end{align}

We apply the Bond transformations (equations \ref{eq:M_rot},  \ref{eq:Psi_rot} and \ref{eq:X_rot}) to equations \ref{eq:M2ndgen_newsys} through \ref{eq:X2ndgen_newsys} and compare the result with equations
\ref{eq:M2ndgen_newsys} through \ref{eq:X2ndgen_newsys}. Finally, the transformation relations for the coefficient matrices are given by
\begin{align} \label{eq:Mn_rot}
& \mathbf{M}^{(n)} = \mathbf{L} \tilde{\mathbf{M}}^{(n)} \mathbf{L}^T , \quad n=0,1,2, \\
& \mathbf{K}^{(m)} = \mathbf{L} \tilde{\mathbf{K}}^{(m)} \mathbf{R}^T , \quad m=1,2, \\
& \mathbf{J}^{(0)} = \mathbf{R} \tilde{\mathbf{J}}^{(0)} \mathbf{R}^{T} .
\end{align}

\subsection{A numerical comparison with the Kolsky and Kjartansson models}
We design a numerical experiment to compare the first- and second-order nearly constant $Q$ models with the Kolsky and Kjartansson models. We consider the viscoelastic orthorhombic case. The medium density is set as $\rho=10^3$~$\text{kg/m}^3$. The reference stiffness coefficients and quality factors are designed as
\begin{equation}
\mathbf{M}^{(0)} =
\left(
\begin{matrix}
9.00 & 3.60 & 2.25 & 0    & 0    & 0    \\
     & 9.84 & 2.40 & 0    & 0    & 0    \\
     &      & 5.94 & 0    & 0    & 0    \\
     &      &      & 2.00 & 0    & 0    \\
     &      &      &      & 1.60 & 0    \\
     &      &      &      &      & 2.18
\end{matrix}
\right) ,
\mathbf{Q} =
\left(
\begin{matrix}
70   &   35 &   45 & \infty  & \infty  & \infty    \\
     &   60 &   48 & \infty  & \infty  & \infty    \\
     &      &   50 & \infty  & \infty  & \infty    \\
     &      &      & 35      & \infty  & \infty    \\
     &      &      &         & 30      & \infty    \\
     &      &      &         &         & 40
\end{matrix}
\right) ,
\end{equation}
where $\mathbf{M}^{(0)}$ and $\mathbf{Q}$ are symmetric matrices and only their upper diagonal elements are shown for brevity. The units of $\mathbf{M}^{(0)}$ are $10^9$~Pa. The substance with material properties $\mathbf{M}^{(0)}$ was proposed by \cite{schoenberg:1997}. 

Matrices $\mathbf{M}^{(0)}$ and $\mathbf{Q}$ are used to initialize the complex stiffness coefficients for the first- and second-order nearly constant $Q$ models (equations \ref{eq:M1stgen} and \ref{eq:M2ndgen} together with equations \ref{eq:MIJn}) and the Kjartansson and Kolsky models (equation \ref{eq:Mkjar} together with equation \ref{eq:gamma_kjar}, and equation \ref{eq:Mkolsky}). Table \ref{tab:tabl1} shows the relaxation times in the weighing function \ref{eq:Weq} for the first- and second-order nearly constant $Q$ models. The reference frequency is set as $f_{0} = 100$~Hz for all the dissipative models. 

\begin{table}[!h]
\vspace{-2ex}
\centering
\caption{Optimal relaxation times for the five-element weighting function in the frequency range $[1,200]$~Hz, from Table 11 of \cite{hao.greenhalgh:2021}.}
\label{tab:tabl1}
\begin{tabular}{c c c}
\toprule
$\ell$ & $\tau_{\sigma}^{(\ell)}$ (s) & $\Delta \tau^{(\ell)} = \tau_{\epsilon}^{(\ell)}-\tau_{\sigma}^{(\ell)}$ (s) \\
\midrule
1 &	1.8230838 $\times 10^{-1}$ & 2.7518001 $\times 10^{-1}$ \\
2 &	3.2947348 $\times 10^{-2}$ & 3.0329269 $\times 10^{-2}$ \\
3 &	8.4325390 $\times 10^{-3}$ & 6.9820198 $\times 10^{-3}$ \\
4 &	2.3560480 $\times 10^{-3}$ & 1.9223614 $\times 10^{-3}$ \\
5 &	5.1033826 $\times 10^{-4}$ & 7.2390630 $\times 10^{-4}$ \\
\bottomrule     
\end{tabular}
\vspace{-2ex}
\end{table}

We use the following formulas to compute the quality factor and phase velocity of a homogeneous plane wave
\begin{equation} \label{eq:QV}
Q = -\frac{\text{Re}(v^2)}{\text{Im}(v^2)} , \qquad
V = \frac{v_{R}^2 + v_{I}^2}{v_{R}} .
\end{equation}
Here, $v=v_{R}-i \text{sgn}(\omega) v_{I}$ denotes the complex velocity, the square of which is an eigenvalue of the Christoffel matrix \cite{cerveny:2001}. The minus sign corresponds to the Fourier transform definition (equation \ref{eq:Fourier}). Quantities $v_{R}$ and $v_{I}$ are real valued. The above definitions for $Q$ and $V$ were given by \cite{carcione:1995} and \cite{knopoff:1964}, respectively. The supplementary material provided shows the Christoffel matrix and the stiffness matrices for all the dissipative models in the orthorhombic case.  

\begin{figure}[hbt!]
\centering
\subfloat[$(0^{\degree},0^{\degree})$]{\includegraphics[width=4.5cm]{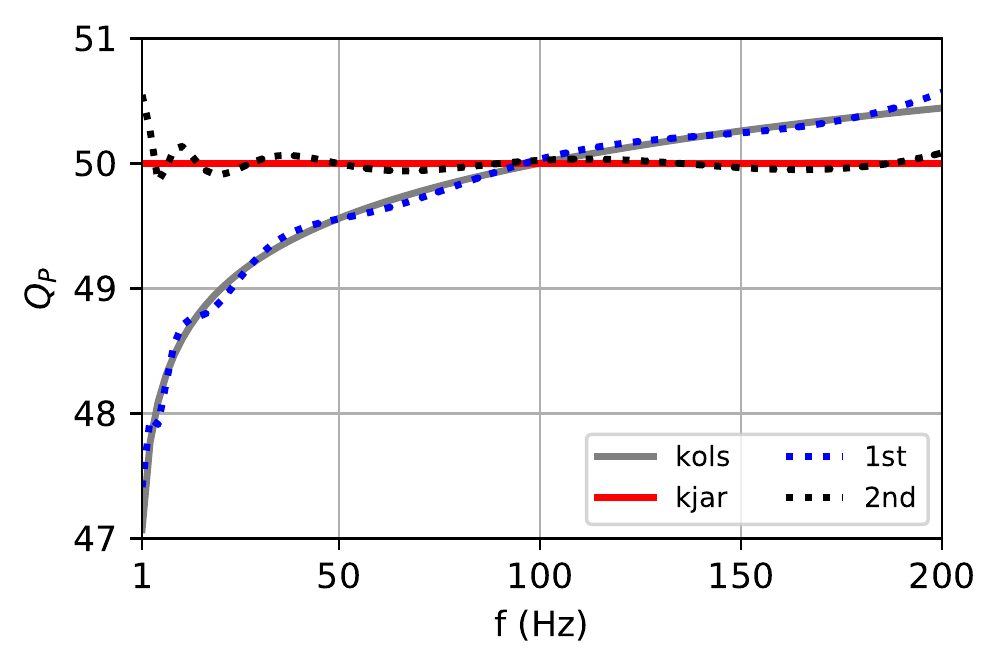}
\label{fig:Qp_ang0}} 
\subfloat[$(90^{\degree},0^{\degree})$]{\includegraphics[width=4.5cm]{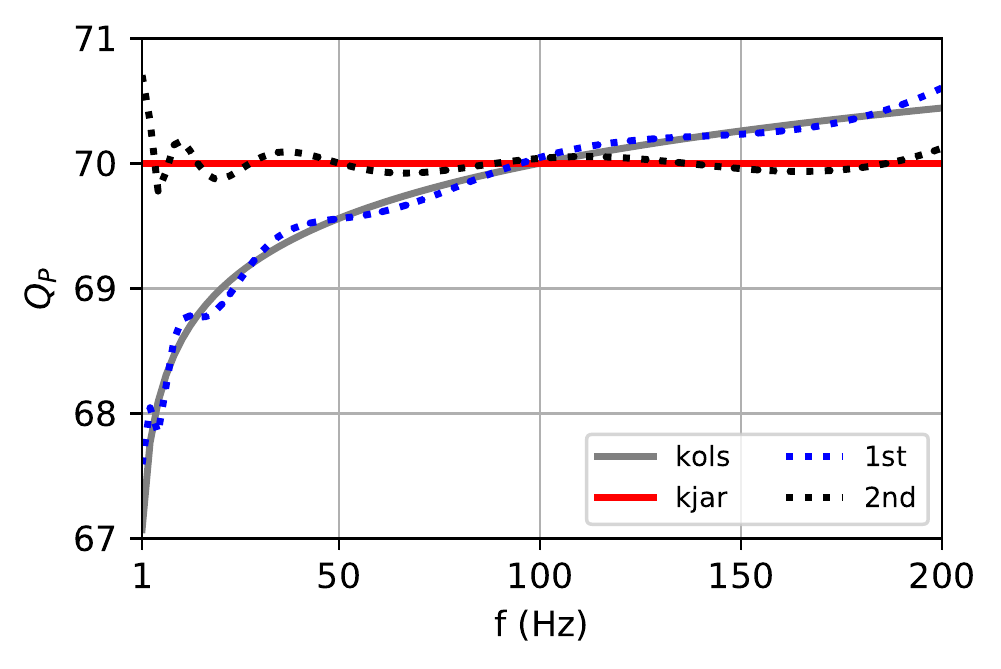}
\label{fig:Qp_ang1}}
\subfloat[$(90^{\degree},90^{\degree})$]{\includegraphics[width=4.5cm]{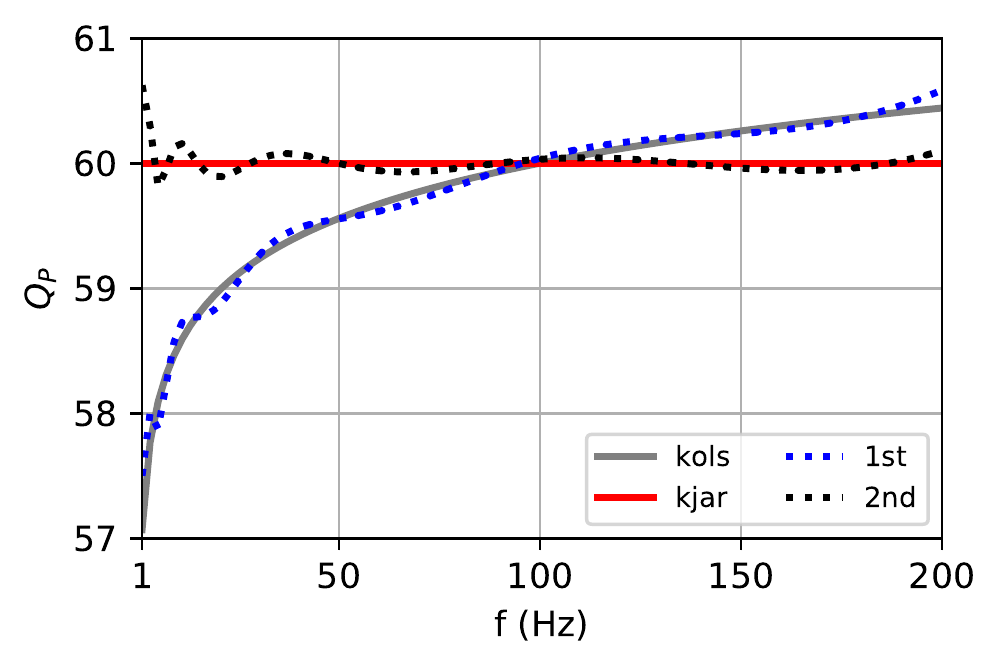}
\label{fig:Qp_ang2}}
\caption{
The quality factors of homogeneous plane P waves in dissipative models for orthorhombic anisotropy. The abbreviations ``Kols'' and ``Kjar'' denote the Kolsky and Kjartansson models. The abbreviations ``1st'' and ``2nd'' denote the first- and second-order nearly constant $Q$ models. Quantity $(\theta, \phi)$ describes the wave propagation direction, where $\theta$ denotes the polar angle measured from the $z$ axis, and $\phi$ denotes the azimuthal angle measured in the [x, y] plane and counterclockwise from the $x$ axis. In this manner, $(0^{\degree},0^{\degree})$, $(90^{\degree},0^{\degree})$ and $(90^{\degree},90^{\degree})$ denote the $z$, $x$ and $y$ directions, respectively.  
}
\label{fig:Qp}
\end{figure}

\begin{figure}[hbt!]
\centering
\subfloat[$(0^{\degree},0^{\degree})$]{\includegraphics[width=4.5cm]{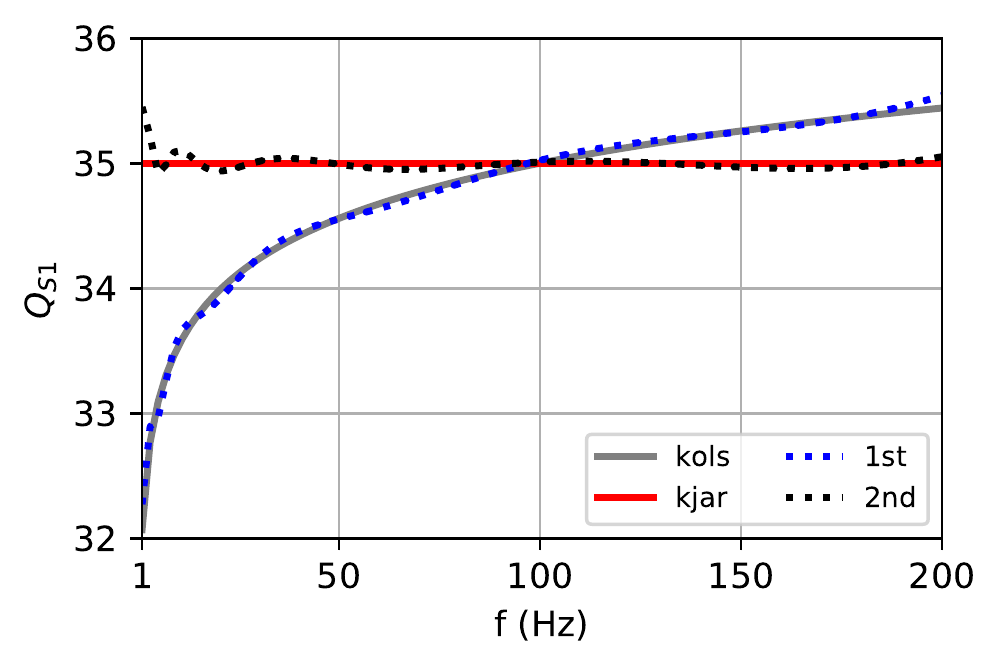}
\label{fig:Qs1_ang0}} 
\subfloat[$(90^{\degree},0^{\degree})$]{\includegraphics[width=4.5cm]{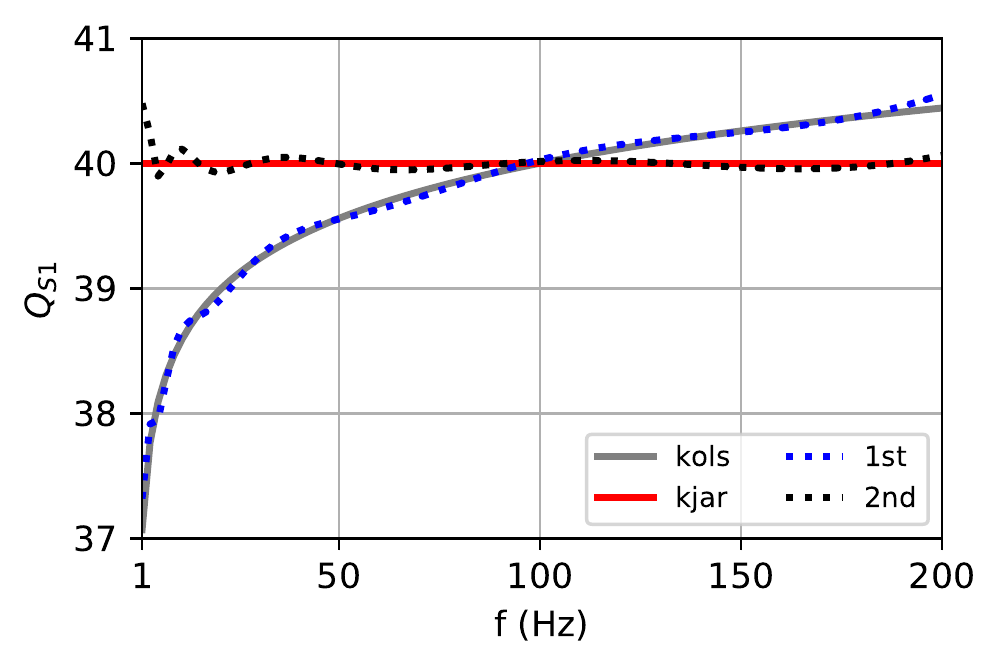}
\label{fig:Qs1_ang1}}
\subfloat[$(90^{\degree},90^{\degree})$]{\includegraphics[width=4.5cm]{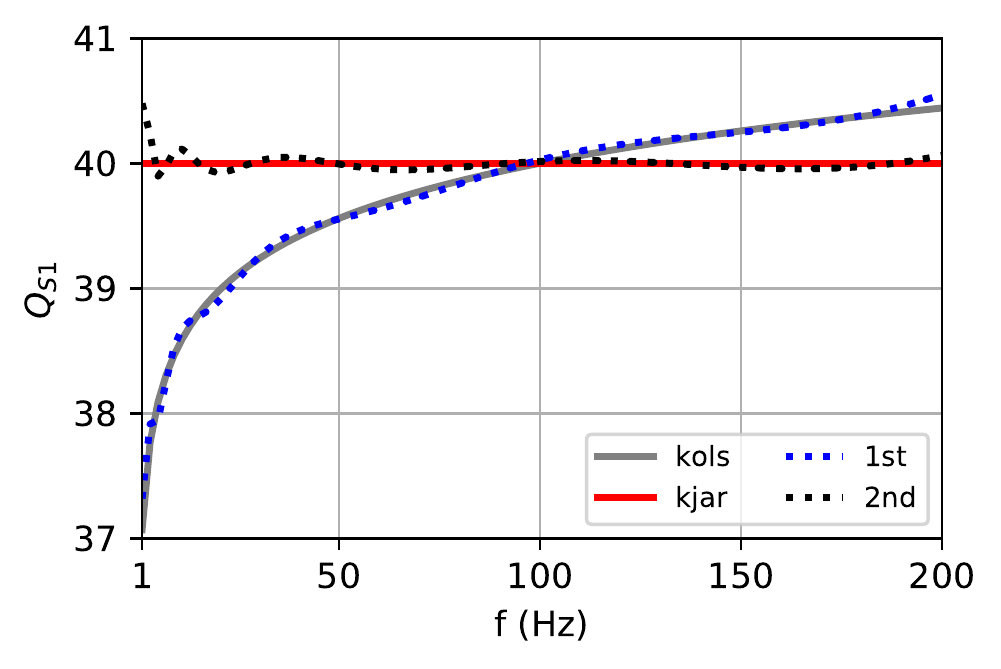}
\label{fig:Qs1_ang2}}
\caption{
Similar to Figure \ref{fig:Qp} but for S$_{1}$ waves.  
}
\label{fig:Qs1}
\end{figure}

\begin{figure}[hbt!]
\centering
\subfloat[$(0^{\degree},0^{\degree})$]{\includegraphics[width=4.5cm]{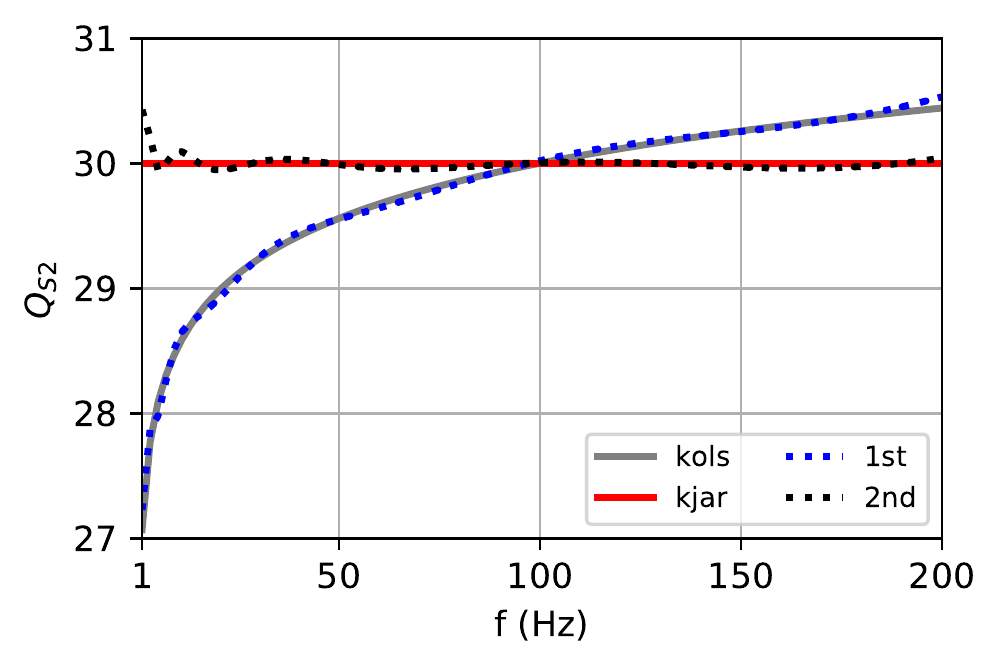}
\label{fig:Qs2_ang0}} 
\subfloat[$(90^{\degree},0^{\degree})$]{\includegraphics[width=4.5cm]{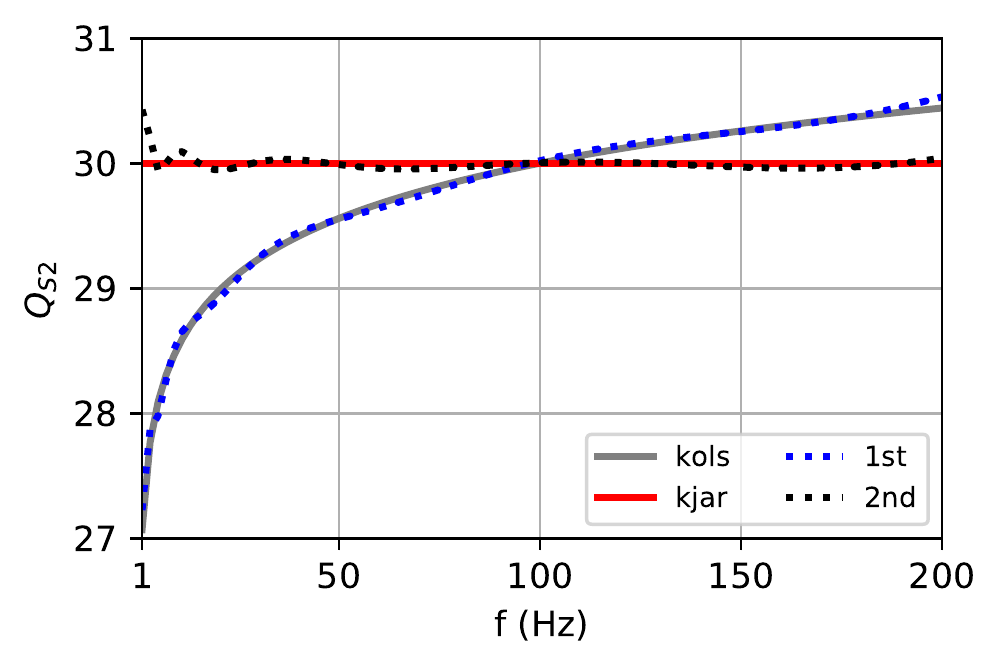}
\label{fig:Qs2_ang1}}
\subfloat[$(90^{\degree},90^{\degree})$]{\includegraphics[width=4.5cm]{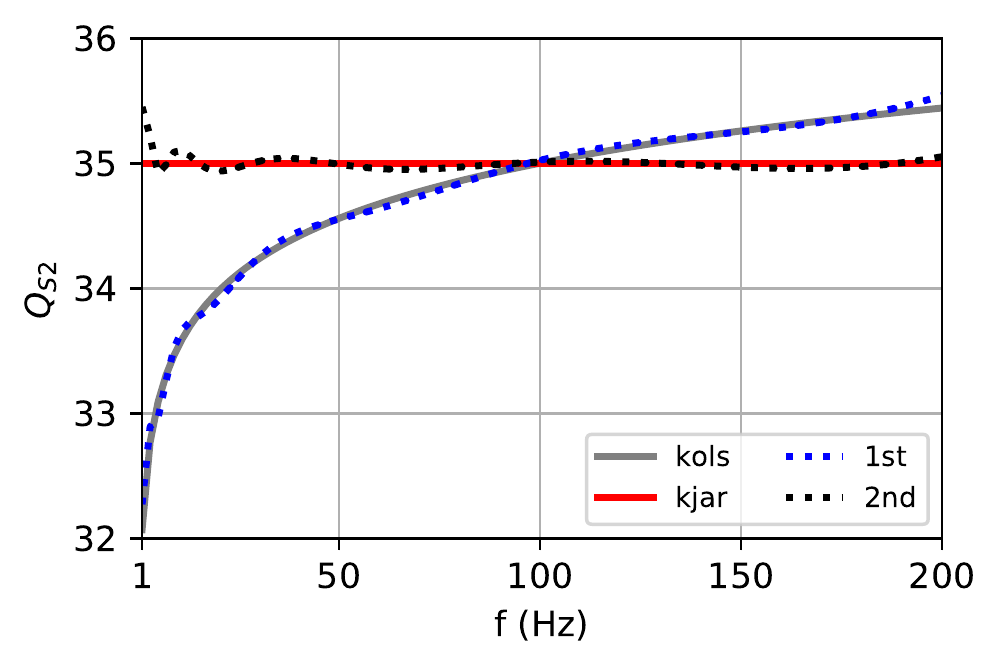}
\label{fig:Qs2_ang2}}
\caption{
Similar to Figure \ref{fig:Qp} but for S$_{2}$ waves.  
}
\label{fig:Qs2}
\end{figure}

\begin{figure}[hbt!]
\centering
\subfloat[$(0^{\degree},0^{\degree})$]{\includegraphics[width=4.5cm]{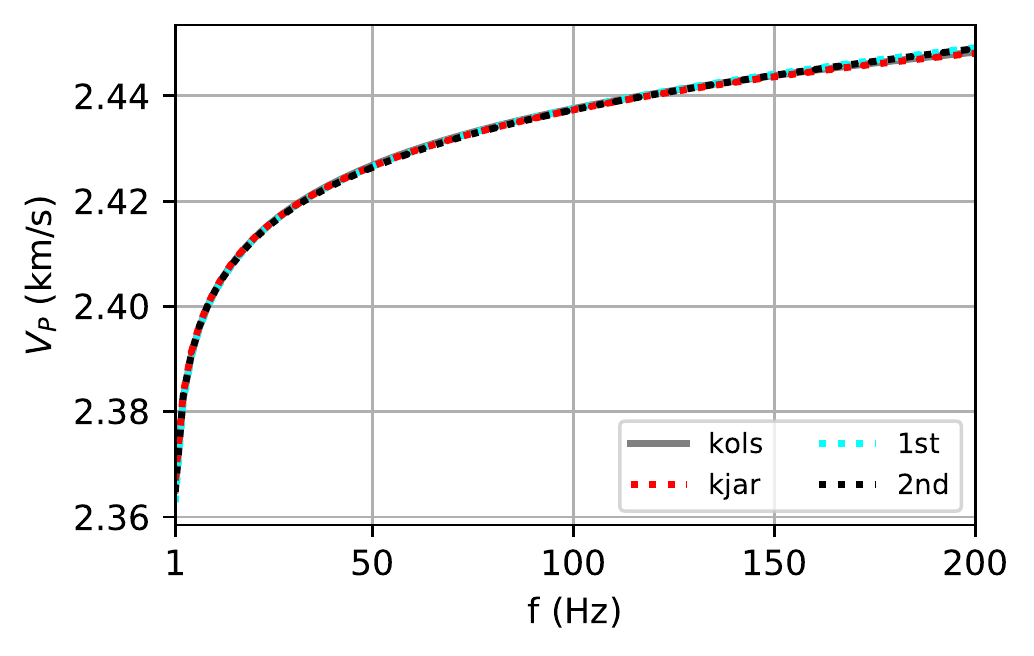}
\label{fig:Vp_ang0}} 
\subfloat[$(90^{\degree},0^{\degree})$]{\includegraphics[width=4.5cm]{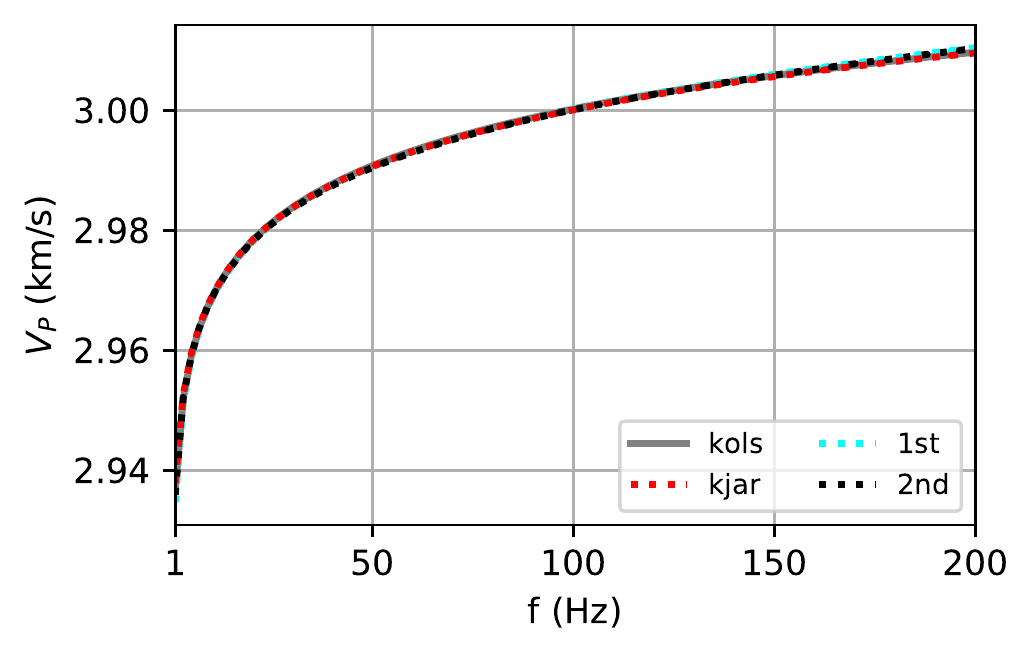}
\label{fig:Vp_ang1}}
\subfloat[$(90^{\degree},90^{\degree})$]{\includegraphics[width=4.5cm]{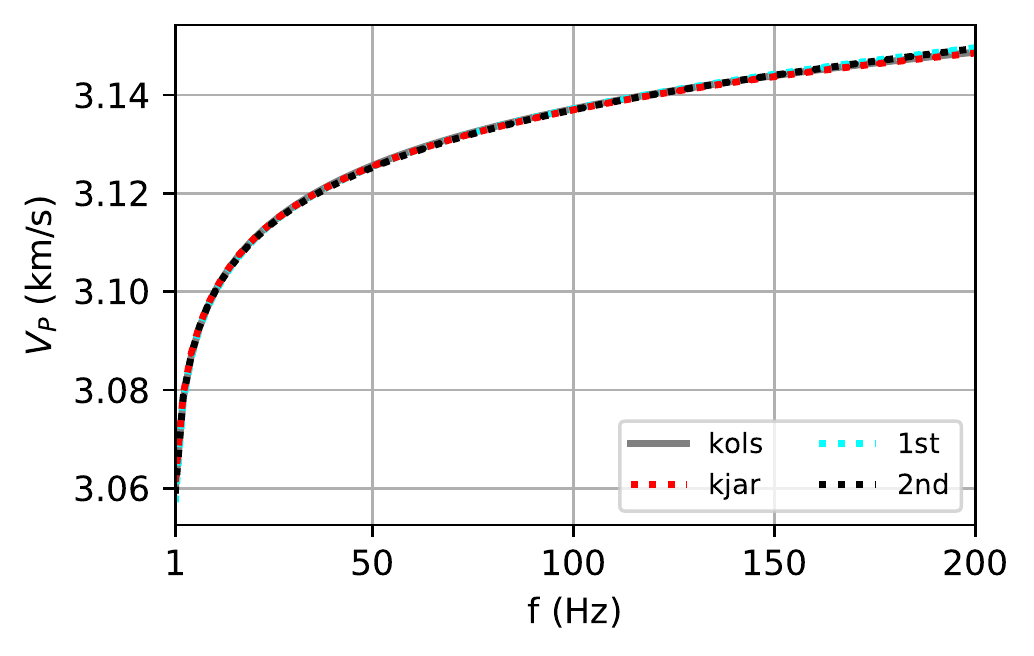}
\label{fig:Vp_ang2}}
\caption{
Similar to Figure \ref{fig:Qp} but for the phase velocities of homogeneous plane P waves. 
}
\label{fig:Vp}
\end{figure}

\begin{figure}[hbt!]
\centering
\subfloat[$(0^{\degree},0^{\degree})$]{\includegraphics[width=4.5cm]{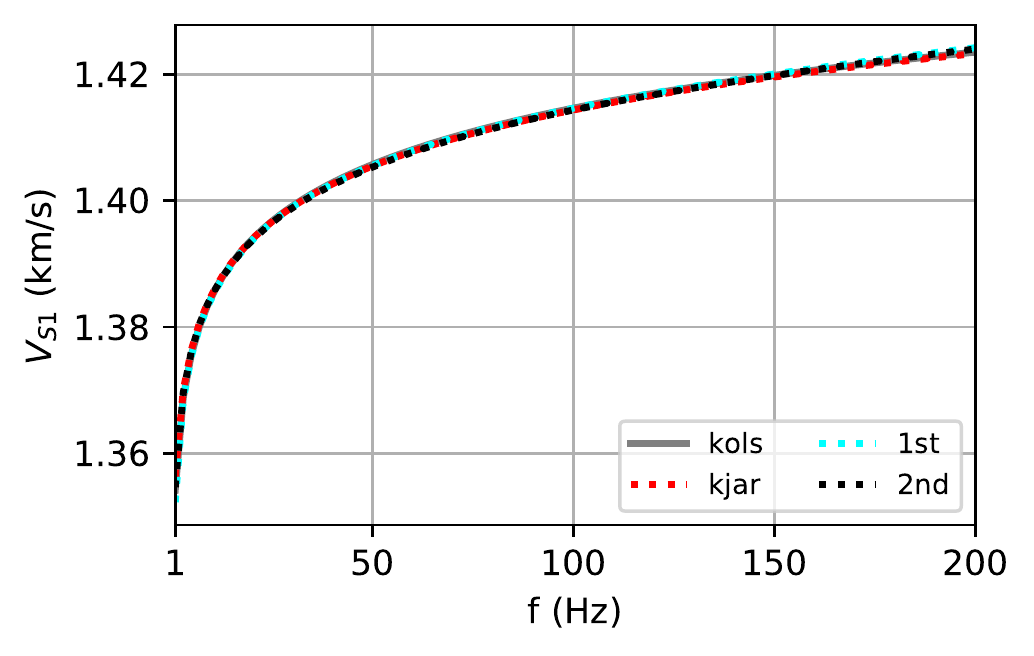}
\label{fig:Vs1_ang0}} 
\subfloat[$(90^{\degree},0^{\degree})$]{\includegraphics[width=4.5cm]{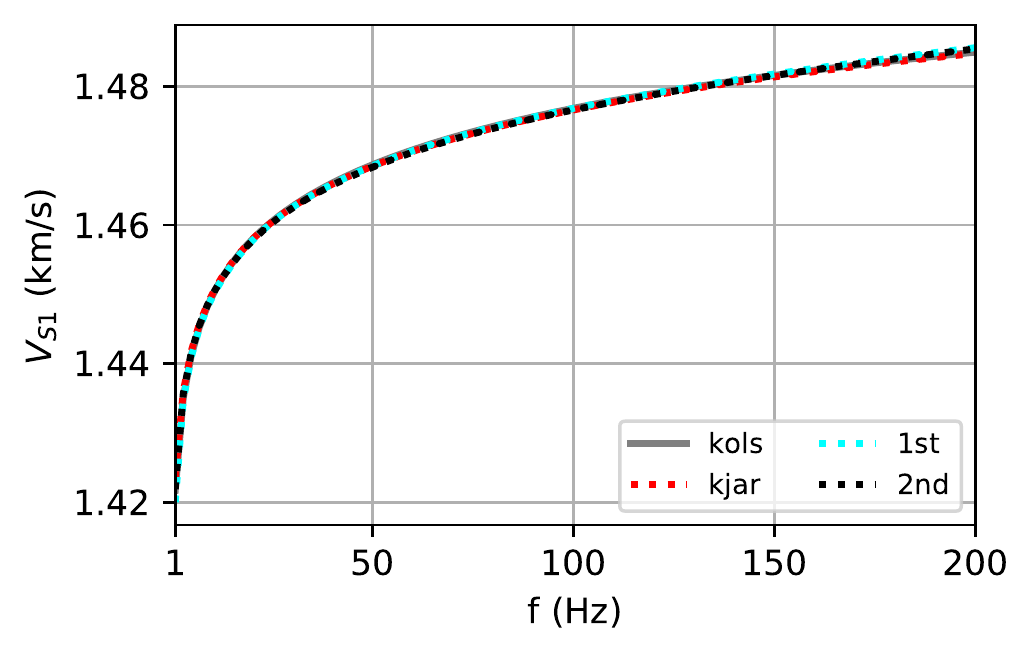}
\label{fig:Vs1_ang1}}
\subfloat[$(90^{\degree},90^{\degree})$]{\includegraphics[width=4.5cm]{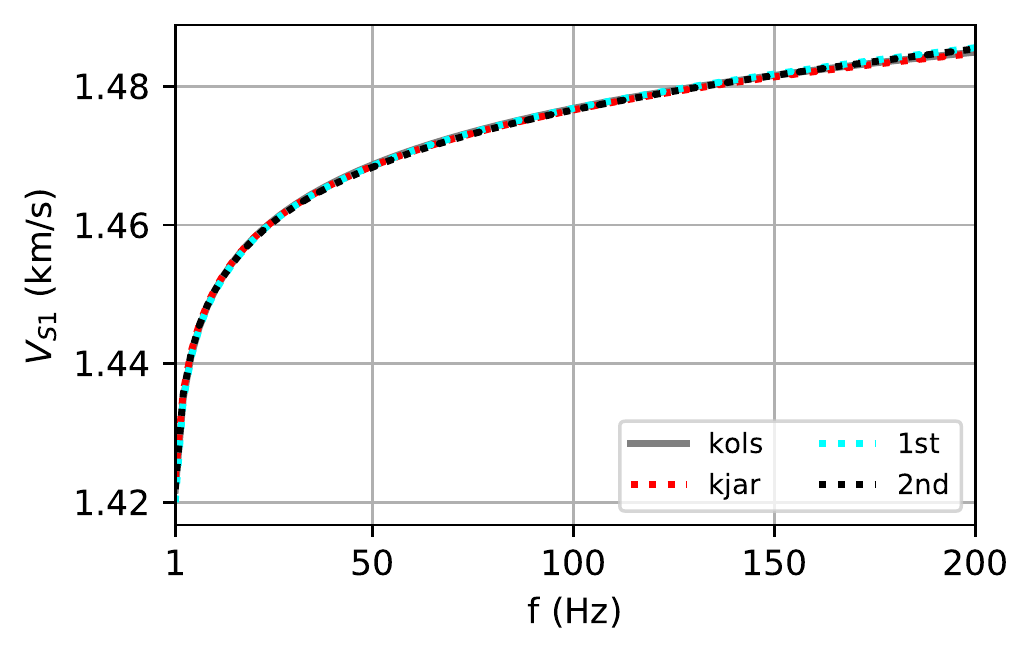}
\label{fig:Vs1_ang2}}
\caption{
Similar to Figure \ref{fig:Vp} but for S$_{1}$ waves.  
}
\label{fig:Vs1}
\end{figure}

\begin{figure}[hbt!]
\centering
\subfloat[$(0^{\degree},0^{\degree})$]{\includegraphics[width=4.5cm]{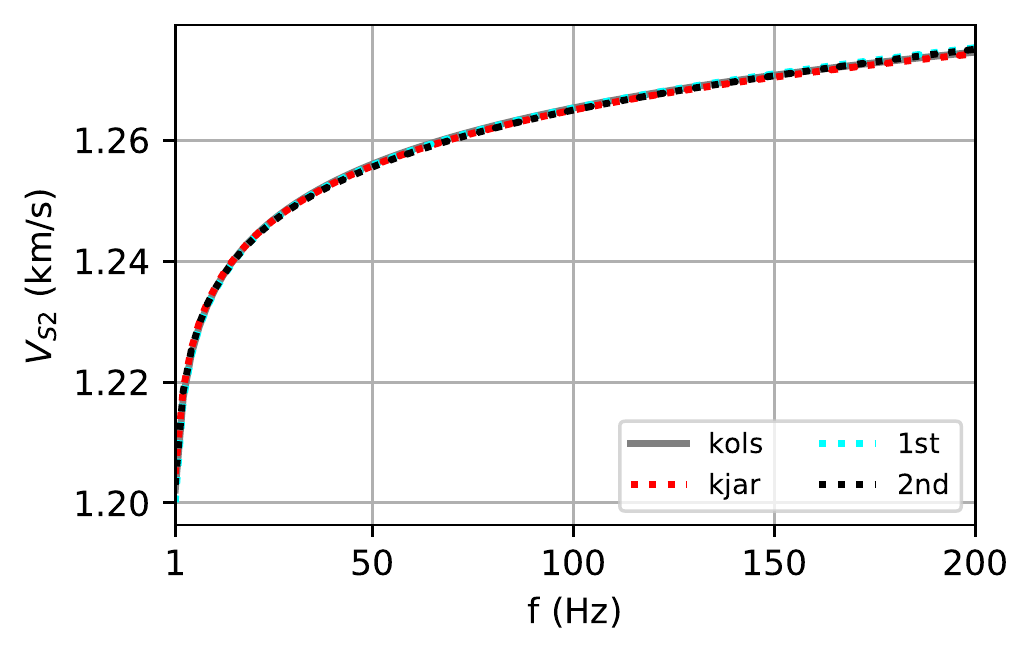}
\label{fig:Vs2_ang0}} 
\subfloat[$(90^{\degree},0^{\degree})$]{\includegraphics[width=4.5cm]{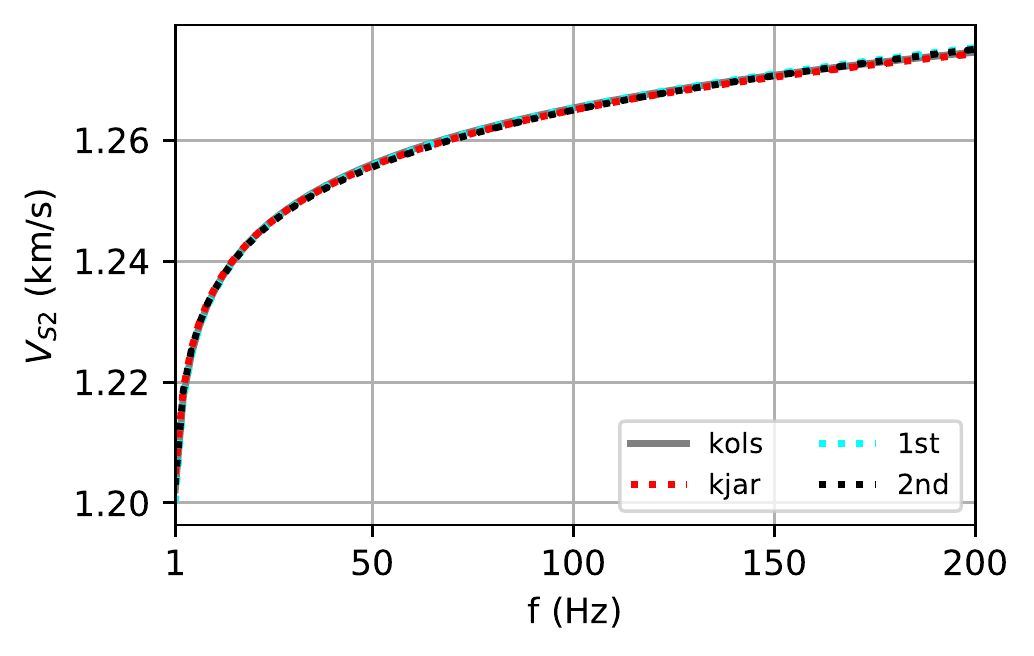}
\label{fig:Vs2_ang1}}
\subfloat[$(90^{\degree},90^{\degree})$]{\includegraphics[width=4.5cm]{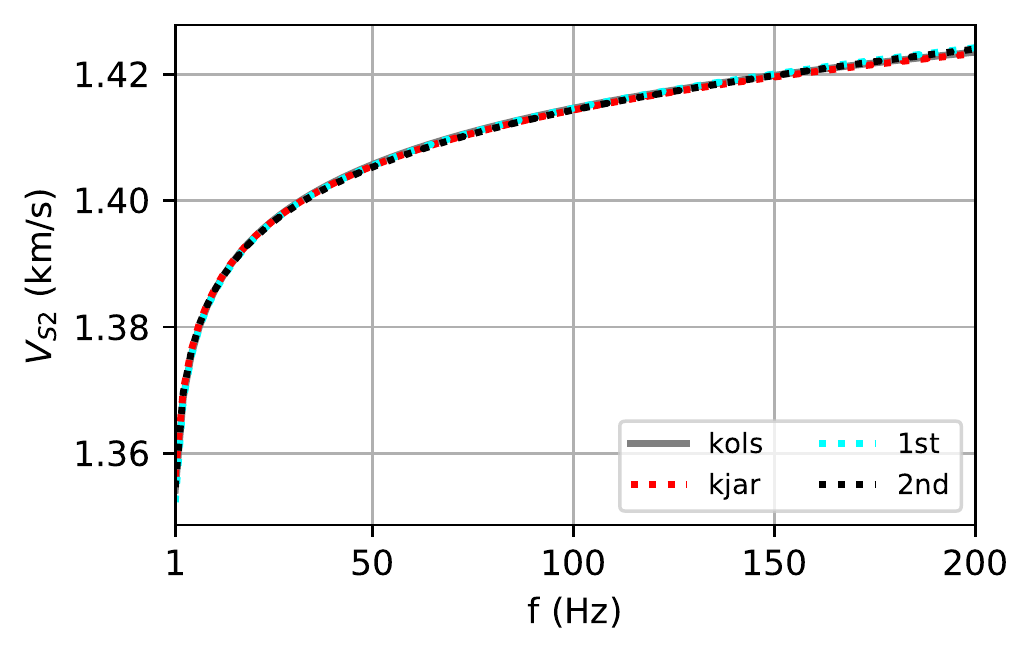}
\label{fig:Vs2_ang2}} 
\caption{
Similar to Figure \ref{fig:Vp} but for S$_{2}$ waves.  
}
\label{fig:Vs2}
\end{figure}

We take account of homogeneous plane P, S$_{1}$ and S$_{2}$ waves, where P denotes the quasi-primary wave, and S$_{1}$ and S$_{2}$ denote the fast and slow quasi-shear waves, respectively. For a specified propagation direction, $v_{R}$ of an S$_{1}$ wave is larger than that of an S$_{2}$ wave. Figures \ref{fig:Qp} through \ref{fig:Qs2} indicate that the quality factors for the first- and second-order order nearly constant $Q$ model are accurate approximations to those for the Kolsky and Kjartansson models, respectively. The second-order nearly constant $Q$ model is closer to being constant $Q$ than the first-order one. All the figures suggest that the Kjartansson model is not exactly constant $Q$ in the orthorhombic case. Figures \ref{fig:Vp} through \ref{fig:Vs2} show that the corresponding phase velocities for all three wave types (P, S$_{1}$ and S$_{2}$) for the different models and different propagation directions. For any of the three wave types, the phase velocities from the first- and second-order nearly constant $Q$ models fit well with those from the Kolsky and Kjartansson models. The supplementary material file shows the figures for the quality factor and phase velocity in more propagation directions, and the variation of the anisotropy parameters with frequency.  

\section{Nearly constant $Q$ wave equations}
In this section, we derive the wave equations in differential form for the first- and second-order nearly constant $Q$ models. For convenience, we adopt the tensor notation to describe the relaxation functions and wave equations. The relaxation functions, the creep functions, the complex stiffness coefficients and the complex compliances individually satisfy the correspondence relationship between the fourth-rank tensor notation and the two-index Voigt notation. For example, quantities $\psi_{ijkl}$ and $\Psi_{IJ}$ denote the same relaxation functions in the fourth-rank tensor form and two-index Voigt form, respectively. Here, $\Psi_{IJ}$ are the elements of relaxation function matrix $\bm{\Psi}$. The index correspondence relations show that $\psi_{ijkl}$ is identical to $\Psi_{IJ}$ in the following way: $ij$ corresponds to $I$ and $kl$ corresponds to $J$, and correspondence $ij \to I$ and $kl \to J$ is given by: $11\to1$, $22\to2$, $33\to3$, $23\to4$, $13\to5$ and $12\to6$. Besides, tensor $\psi_{ijkl}$ satisfies the symmetry relations, namely $\psi_{ijkl} = \psi_{jikl} = \psi_{ijlk} = \psi_{klij}$.

The wave equations for a general viscoelastic anisotropic medium are given by
\begin{align} 
\label{eq:motion}
& \rho \diffp{{u_{i}}}{{t^2}} = \diffp{{\sigma_{ij}}}{{x_j}} + S_{i} , \\
\label{eq:const_4rnk_gen}
& \sigma_{ij} = \psi_{ijkl} \odot \epsilon_{kl} , \\
\label{eq:geom_eq}
& \epsilon_{kl} = \frac{1}{2} \left(
\diffp{{u_{k}}}{{x_{l}}} + \diffp{{u_{l}}}{{x_{k}}}
\right) ,
\end{align}
where $\rho$ denotes density and $u_{i}$ denotes particle displacement components. Quantity $\sigma_{ij}$ and $\epsilon_{ij}$ denotes the second-rank tensors of stress and strain, respectively. Quantity $S_{i}$ denotes the components of a vector (directed) source function. Equations \ref{eq:motion}, \ref{eq:const_4rnk_gen} and \ref{eq:geom_eq} are the equation of motion, the constitutive equation and the relationship between strain and particle displacement, respectively. The repeated indices in equations \ref{eq:motion} and \ref{eq:const_4rnk_gen} satisfy the Einstein summation convention.   

We next derive the wave equations in differential form for the first- and second-order nearly constant $Q$ models.

\subsection{Wave equations for the first-order nearly constant $Q$ model}
Substituting equation \ref{eq:zeta_new} into equation \ref{eq:Psi1stgen}, the fourth-rank tensor of the relaxation functions for the first-order nearly constant $Q$ model is written as
\begin{equation} \label{eq:psi_1stnew}
\psi_{ijkl} = A_{ijkl}^{(0)} H(t) - A_{ijkl}^{(1)} \sum_{\ell=1}^{L}\xi^{(\ell)}(t) ,
\end{equation}
where $A_{ijkl}^{(0)}$ and $A_{ijkl}^{(1)}$ are given by:
\begin{align}
& A_{ijkl}^{(0)} = M_{ijkl}^{(0)} + g M_{ijkl}^{(1)} , \\
& A_{ijkl}^{(1)} = M_{ijkl}^{(1)} .
\end{align}
Here, $M_{ijkl}^{(n)}$, $n=0,1,2$, is the tensor form of matrix $\mathbf{M}^{(n)}$. Quantity $g$ is given in equation \ref{eq:gnew}.

Substitution of equation \ref{eq:psi_1stnew} into equation \ref{eq:const_4rnk_gen} and utilizing property \ref{eq:xil_odot_eps} results in the constitutive relation for the first-order nearly constant $Q$ model
\begin{align}
\label{eq:sigma1st}
& \sigma_{ij} = A_{ijkl}^{(0)}\epsilon_{kl} - \sum_{\ell=1}^{L} \alpha_{ij}^{(\ell)} , \\
\label{eq:q1st}
& \alpha_{ij}^{(\ell)} = \dot{\xi}^{(\ell)} * \left(A_{ijkl}^{(1)}\epsilon_{kl}\right) ,
\end{align}

Referring to equation \ref{eq:zetaxiprop}, equation \ref{eq:q1st} is written in differential form as
\begin{equation} \label{eq:dqdt1st}
\diffp{{\alpha_{ij}^{(\ell)}}}{t} = s^{(\ell)}  
A_{ijkl}^{(1)} \epsilon_{kl} - \frac{1}{\tau_{\sigma}^{(\ell)}} \alpha_{ij}^{(\ell)} ,
\end{equation}
where $s^{(\ell)}$ is given in equation \ref{eq:sl}.

We substitute equation \ref{eq:geom_eq} into equation \ref{eq:sigma1st} and then substitute the result into equation \ref{eq:motion} and take the divergence of equation \ref{eq:dqdt1st}. Finally, the wave equations for the first-order nearly constant $Q$ model are written in terms of the particle displacement as
\begin{align}
& \rho \diffp{{u_{i}}}{{t^2}} = \diffp{}{{x_{j}}} 
\left( A_{ijkl}^{(0)} \diffp{{u_{k}}}{{x_l}} \right)
- \sum_{\ell=1}^{L} r_{i}^{(\ell)} + S_{i} , \\
& \diffp{{r_{i}^{(\ell)}}}{t} = s^{(\ell)} 
 \diffp{}{{x_{j}}} \left( A_{ijkl}^{(1)} \diffp{{u_{k}}}{{x_l}} \right)
- \frac{1}{\tau_{\sigma}^{(\ell)}} r_{i}^{(\ell)} ,
\end{align}
where $r_{i}^{(\ell)} = \partial a_{ij}^{(\ell)}/\partial x_{j} $ are memory variables. 

As an alternative, we substitute equation \ref{eq:geom_eq} into equations \ref{eq:sigma1st} and \ref{eq:dqdt1st} and take the first temporal derivative of the result. Finally, the wave equations can be written in terms of particle velocity and stress as
\begin{align} 
& \rho \diffp{{v_{i}}}{{t}} = \diffp{{\sigma_{ij}}}{{x_j}} + S_{i} , \\
& \diffp{{\sigma_{ij}}}{t} = A_{ijkl}^{(0)} \diffp{{v_{k}}}{{x_{l}}} - \sum_{\ell=1}^{L} w_{ij}^{(\ell)} , \\
& \diffp{{w_{ij}^{(\ell)}}}{t} = s^{(\ell)}  
A_{ijkl}^{(1)} \diffp{{v_{k}}}{{x_{l}}} - \frac{1}{\tau_{\sigma}^{(\ell)}} w_{ij}^{(\ell)} ,
\end{align}
where $v_{i}=\partial u_{i} /\partial t$ denote the particle velocity components. Quantities $w_{ij}^{(\ell)} = \partial \alpha_{ij}^{(\ell)}/\partial t$ are memory variables.

\subsection{Wave equations for the second-order nearly constant $Q$ model}
Substitution of equation \ref{eq:zeta_new} into equation \ref{eq:Psi2ndgen} and utilizing property \ref{eq:xil_odot_eps} gives rise to the fourth-rank tensor of the relaxation functions for the second-order nearly constant $Q$ model
\begin{equation} \label{eq:psi_2ndnew}
\psi_{ijkl} = B_{ijkl}^{(0)} H(t) - B_{ijkl}^{(1)} \sum_{\ell=1}^{L}\xi^{(\ell)}(t) 
 + B_{ijkl}^{(2)} \left( \sum_{\ell=1}^{L} \dot{\xi}^{(\ell)}(t) \right) *
 \left( \sum_{\ell=1}^{L} \xi^{(\ell)}(t) \right) , 
\end{equation}
where $B_{ijkl}^{(0)}$, $B_{ijkl}^{(1)}$ and $B_{ijkl}^{(2)}$ are given by:
\begin{align}
& B_{ijkl}^{(0)} = M_{ijkl}^{(0)} + g M_{ijkl}^{(1)} + 
\frac{1}{2} g^2 M_{ijkl}^{(2)} , \\
& B_{ijkl}^{(1)} = M_{ijkl}^{(1)} + g M_{ijkl}^{(2)} , \\
& B_{ijkl}^{(2)} = \frac{1}{2} M_{ijkl}^{(2)} .
\end{align}

Substituting equation \ref{eq:psi_2ndnew} into equation \ref{eq:const_4rnk_gen} and utilizing the property \ref{eq:xil_odot_eps}, the constitutive relation for the second-order nearly constant $Q$ model is written as
\begin{align}
& \sigma_{ij} = B_{ijkl}^{(0)}\epsilon_{kl} - \sum_{\ell=1}^{L} \beta_{ij}^{(\ell)} , \\
\label{eq:qij2nd1}
& \beta_{ij}^{(\ell)} = \dot{\xi}^{(\ell)} * \left(
B_{ijkl}^{(1)} \epsilon_{kl} - \sum_{\ell=1}^{L} \tilde{\beta}_{ij}^{(\ell)}
\right) , \\
\label{eq:qij2nd2}
& \tilde{\beta}_{ij}^{(\ell)} = \dot{\xi}^{(\ell)} * 
\left( B_{ijkl}^{(2)} \epsilon_{kl} \right) .
\end{align}

Referring to equation \ref{eq:zetaxiprop}, equations \ref{eq:qij2nd1} and \ref{eq:qij2nd2} are written in differential form as
\begin{align}
& \diffp{{\beta_{ij}^{(\ell)}}}{t} = s^{(\ell)} \left(
B_{ijkl}^{(1)} \epsilon_{kl} - \sum_{\ell=1}^{L} \tilde{\beta}_{ij}^{(\ell)}
\right) - \frac{1}{\tau_{\sigma}^{(\ell)}} \beta_{ij}^{(\ell)} , \\
& \diffp{{\tilde{\beta}_{ij}^{(\ell)}}}{t} = s^{(\ell)} 
B_{ijkl}^{(2)} \epsilon_{kl} - \frac{1}{\tau_{\sigma}^{(\ell)}} \tilde{\beta}_{ij}^{(\ell)} ,
\end{align}
where $s^{(\ell)}$ is given in equation \ref{eq:sl}.

We next use the same method as for the first-order nearly constant $Q$ model to obtain the wave equations for the second-order nearly constant $Q$ model. The wave equations in differential form are summarized below. 

The viscoelastic anisotropic wave equations in terms of particle displacement are given by
\begin{align} 
\label{eq:weq2nd_a1}
& \rho \diffp{{u_{i}}}{{t^2}} = \diffp{}{{x_{j}}} 
\left( B_{ijkl}^{(0)} \diffp{{u_{k}}}{{x_l}} \right)
- \sum_{\ell=1}^{L} h_{i}^{(\ell)} + S_{i} , \\
\label{eq:weq2nd_a2}
& \diffp{{h_{i}^{(\ell)}}}{t} = s^{(\ell)} \left[
 \diffp{}{{x_{j}}} \left( B_{ijkl}^{(1)} \diffp{{u_{k}}}{{x_l}} \right) - \sum_{\ell=1}^{L} \tilde{h}_{i}^{(\ell)}
\right]
- \frac{1}{\tau_{\sigma \ell}} h_{i}^{(\ell)} , \\
\label{eq:weq2nd_a3}
& \diffp{{\tilde{h}_{i}^{(\ell)}}}{t} = s^{(\ell)} 
\diffp{}{{x_{j}}} \left( B_{ijkl}^{(2)} \diffp{{u_{k}}}{{x_l}} \right) 
- \frac{1}{\tau_{\sigma}^{(\ell)}} \tilde{h}_{i}^{(\ell)} ,
\end{align}
where $h_{i}^{(\ell)} = \partial \beta_{ij}^{(\ell)}/\partial x_{j}$ and $\tilde{h}_{i}^{(\ell)} = \partial \tilde{\beta}_{ij}^{(\ell)}/\partial x_{j}$ are memory variables.

The wave equations in terms of particle velocity and stress are given by
\begin{align}
& \rho \diffp{{v_{i}}}{{t}} = \diffp{{\sigma_{ij}}}{{x_j}} + S_{i} , \\
& \diffp{{\sigma_{ij}}}{t} = B_{ijkl}^{(0)} \diffp{{v_{k}}}{{x_{l}}} - \sum_{\ell=1}^{L} q_{ij}^{(\ell)} , \\
& \diffp{{q_{ij}^{(\ell)}}}{t} = s^{(\ell)} \left(
B_{ijkl}^{(1)} \diffp{{v_{k}}}{{x_{l}}} - \sum_{\ell=1}^{L} \tilde{q}_{ij}^{(\ell)}
\right)
- \frac{1}{\tau_{\ell}} q_{ij}^{(\ell)} , \\
& \diffp{{\tilde{q}_{ij}^{(\ell)}}}{t} = s^{(\ell)} 
B_{ijkl}^{(2)} \diffp{{v_{k}}}{{x_{l}}} - \frac{1}{\tau_{\sigma}^{(\ell)}} \tilde{q}_{ij}^{(\ell)} ,
\end{align}
where $q_{ij}^{(\ell)} = \partial \beta_{ij}^{(\ell)}/\partial t$ and $\tilde{q}_{ij}^{(\ell)} = \partial \tilde{\beta}_{ij}^{(\ell)}/\partial t$ are memory variables.

\section{Numerical wave modeling in a heterogeneous model}
In this section, we show a numerical example of nearly constant $Q$ wave propagation in the viscoelastic isotropic case. Using equation \ref{eq:Mniso} and the correspondence relationship between the stiffness matrix and the fourth-rank stiffness tensor, the wave equations for the second-order nearly constant $Q$ model (equations \ref{eq:weq2nd_a1} through \ref{eq:weq2nd_a3}) reduce to the following viscoelastic isotropic wave equations (see the supplementary material file)
\begin{align} 
\label{eq:weq_iso1}
& \frac{\partial^{2} \mathbf{u}}{\partial t^{2}} = 
b_{P}^{(0)} \nabla(\nabla \cdot \mathbf{u}) 
- b_{S}^{(0)} \nabla \times \nabla \times \mathbf{u} 
- \sum_{\ell=1}^{L} \mathbf{d}^{(\ell)} + \mathbf{f}, \\
\label{eq:weq_iso2}
& \frac{\partial \mathbf{d}^{(\ell)}}{\partial t} = 
s^{(\ell)} \left( b_{P}^{(1)} \nabla(\nabla \cdot \mathbf{u}) 
- b_{S}^{(1)} \nabla \times \nabla \times \mathbf{u}  
- \sum_{\ell=1}^{L} \tilde{\mathbf{d}}^{(\ell)} \right)
- \frac{1}{\tau_{\sigma}^{(\ell)}} \mathbf{d}^{(\ell)} , \\
\label{eq:weq_iso3}
& \frac{\partial \tilde{\mathbf{d}}^{(\ell)}}{\partial t} = 
s^{(\ell)} \left( b_{P}^{(2)} \nabla(\nabla \cdot \mathbf{u})  
- b_{S}^{(2)} \nabla \times \nabla \times \mathbf{u} \right)
- \frac{1}{\tau_{\sigma}^{(\ell)}} \tilde{\mathbf{d}}^{(\ell)}  ,
\end{align}
where we have assumed the medium to be homogeneous. Quantity $s^{(\ell)}$ is given in equation \ref{eq:sl}. The symbol ``$\nabla$'' denotes the gradient operator. The quantity $\mathbf{u}=(u_{x}, u_{y}, u_{z})^T$ denotes the particle displacement vector. The vector $\mathbf{f}=\mathbf{S}/\rho$ denotes the body force per unit mass. Quantities  $\mathbf{d}^{(\ell)}=\left(d_{x}^{(\ell)},d_{y}^{(\ell)},d_{z}^{(\ell)}\right)^T$ and
$\tilde{\mathbf{d}}^{(\ell)}=\left(\tilde{d}_{x}^{(\ell)},\tilde{d}_{y}^{(\ell)},\tilde{d}_{z}^{(\ell)}\right)^T$ are the density-normalized memory variables in vector form.  
Quantities $b_{I}^{(0)}$, $b_{I}^{(1)}$ and $b_{I}^{(2)}$, $I = P \text{ and } S$, are given by:
\begin{align}
& b_{I}^{(0)} = v_{I}^2 \left(
1 + \frac{g}{Q_{I}} + \frac{g^2}{2Q_{I}^2} \right)  , \\
& b_{I}^{(1)}= \frac{v_{I}^2}{Q_{I}} \left(1 + \frac{g}{Q_{I}} \right), \\
& b_{I}^{(2)} = \frac{v_{I}^2}{2Q_{I}^2} .
\end{align}
Omitting the terms associated with $Q_{I}^2$, the above viscoelastic isotropic wave equations can reduce to the ones for the first-order nearly constant $Q$ model.

\begin{figure}[htb!]
\centering
\subfloat[$v_{P}$]{\includegraphics[width=6cm]{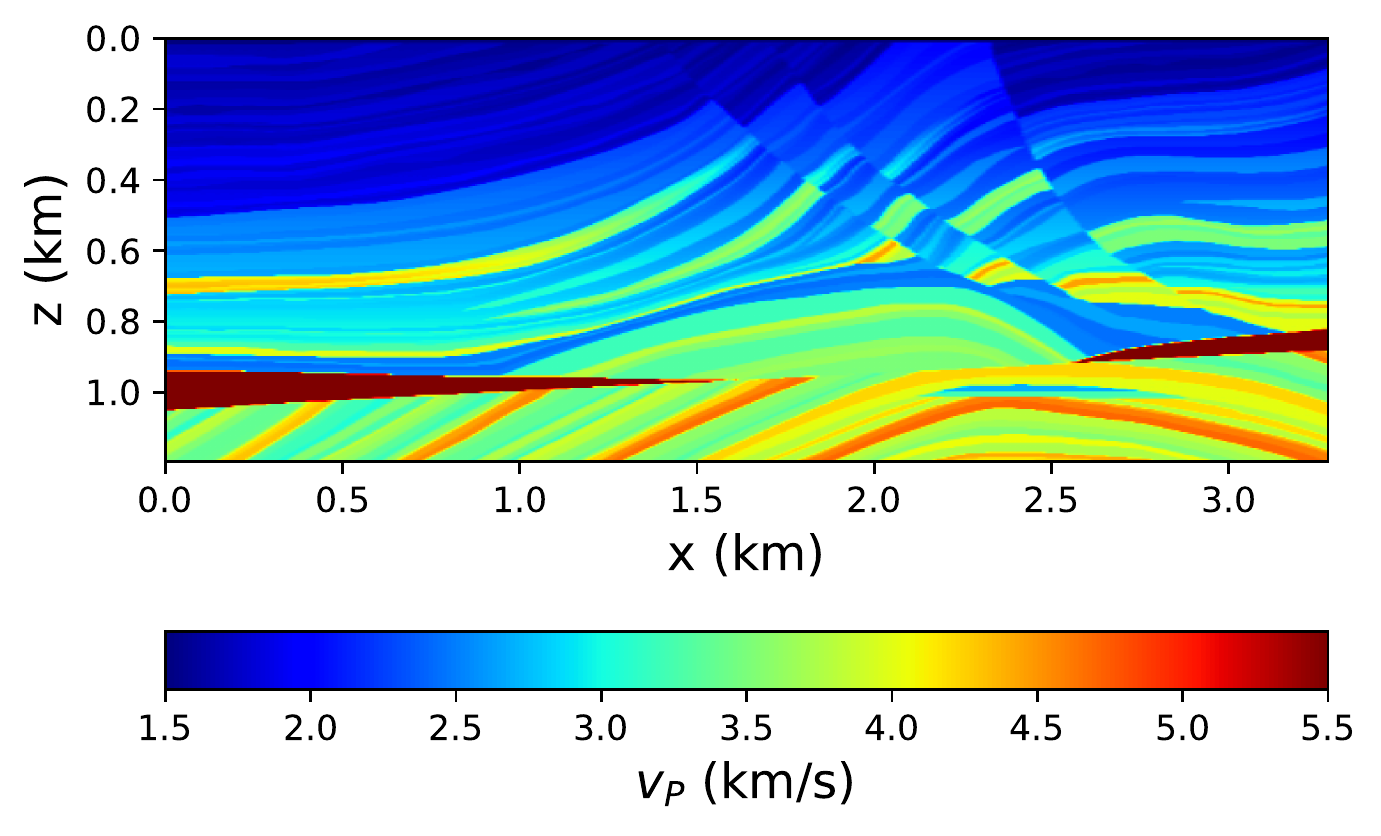}
\label{fig:vp}} 
\quad
\subfloat[$1/Q_{P}$]{\includegraphics[width=6cm]{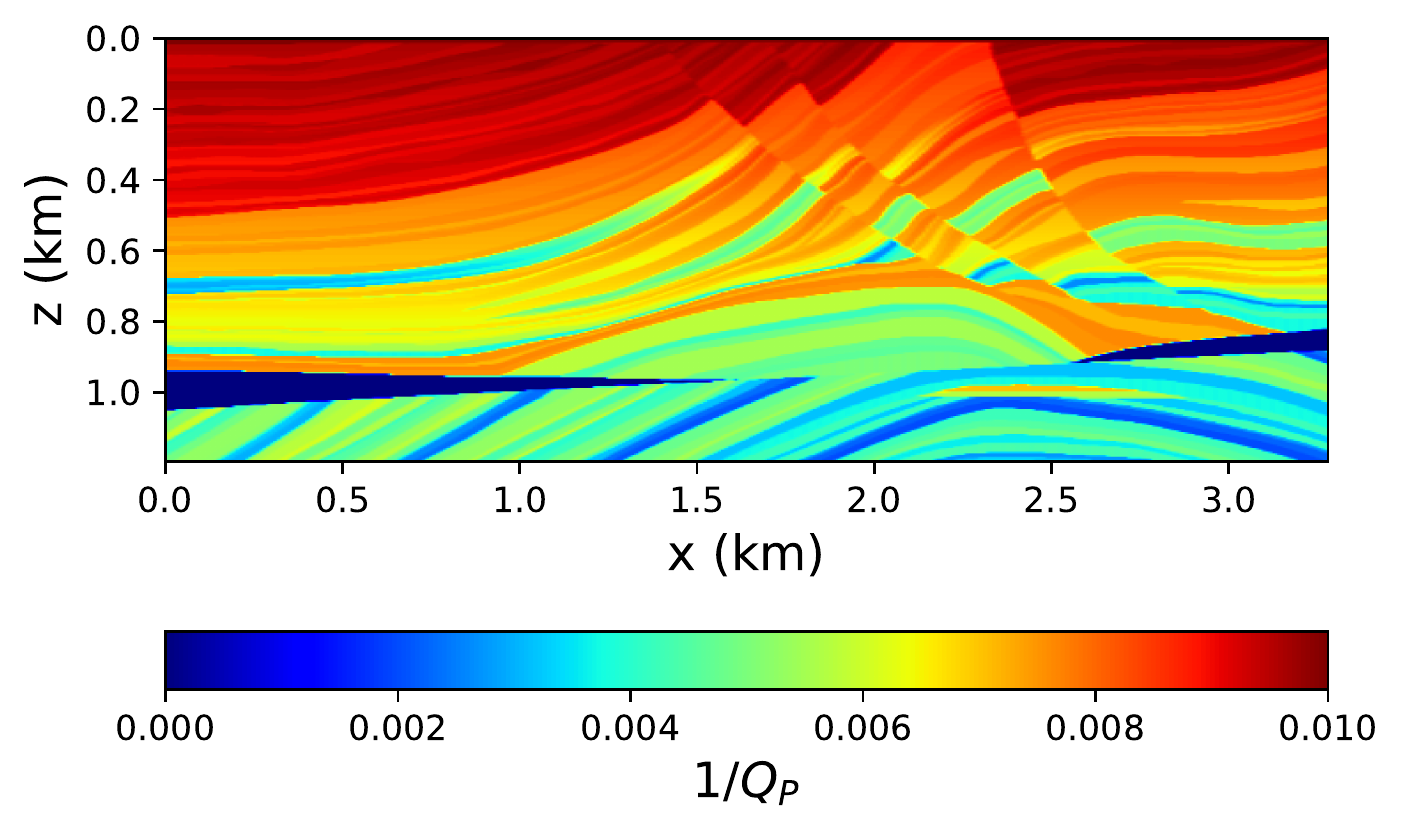}
\label{fig:iQp}} 
\caption{
The P-wave model parameters in the viscoelastic Marmousi model.       
}
\label{fig:marm}
\end{figure}

\begin{figure}[hbt!]
\centering
\subfloat[Elastic and viscoelastic waves]
{\includegraphics[width=7.3cm]{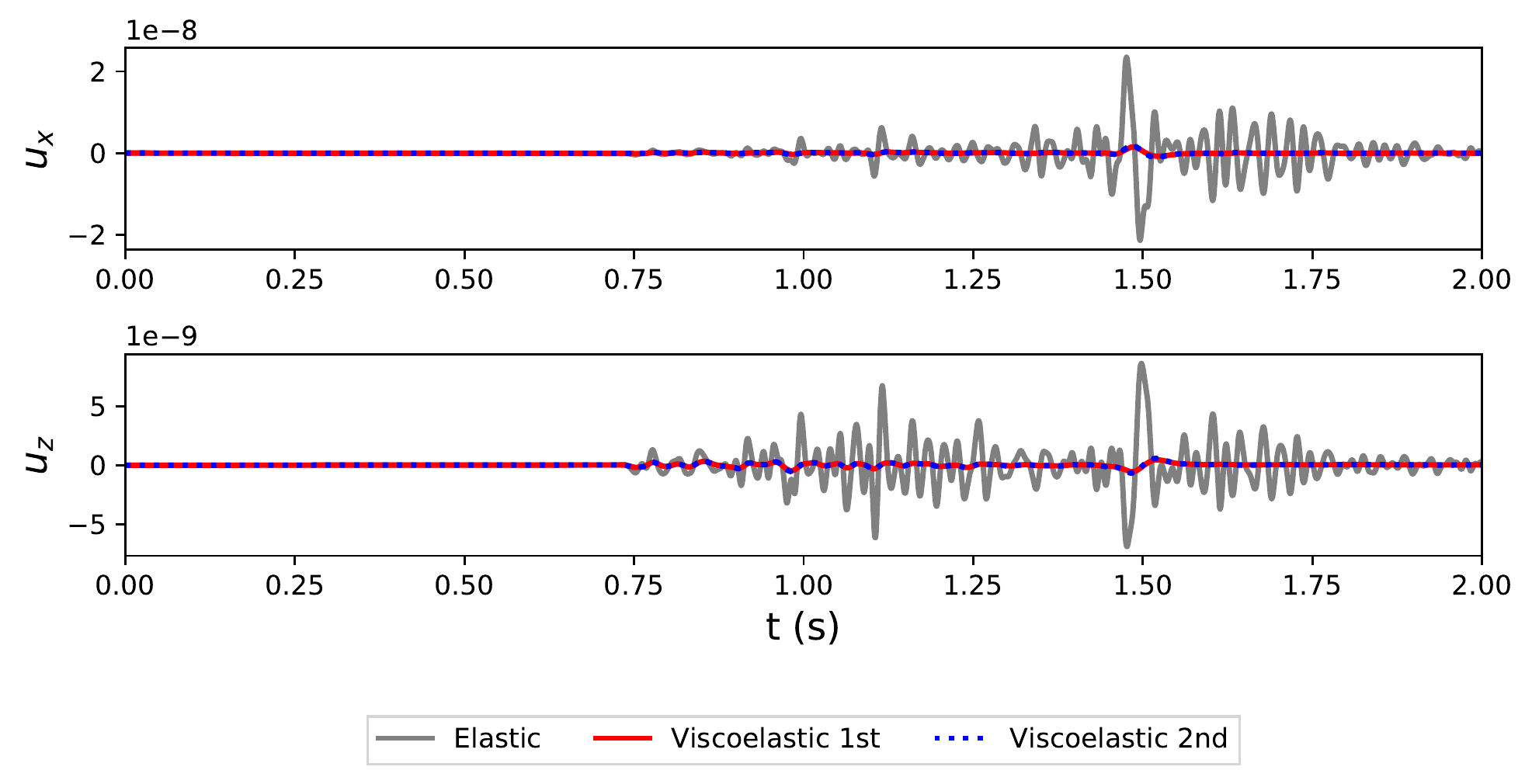}
\label{fig:left1}} 
\subfloat[Viscoelastic waves]
{\includegraphics[width=7.3cm]{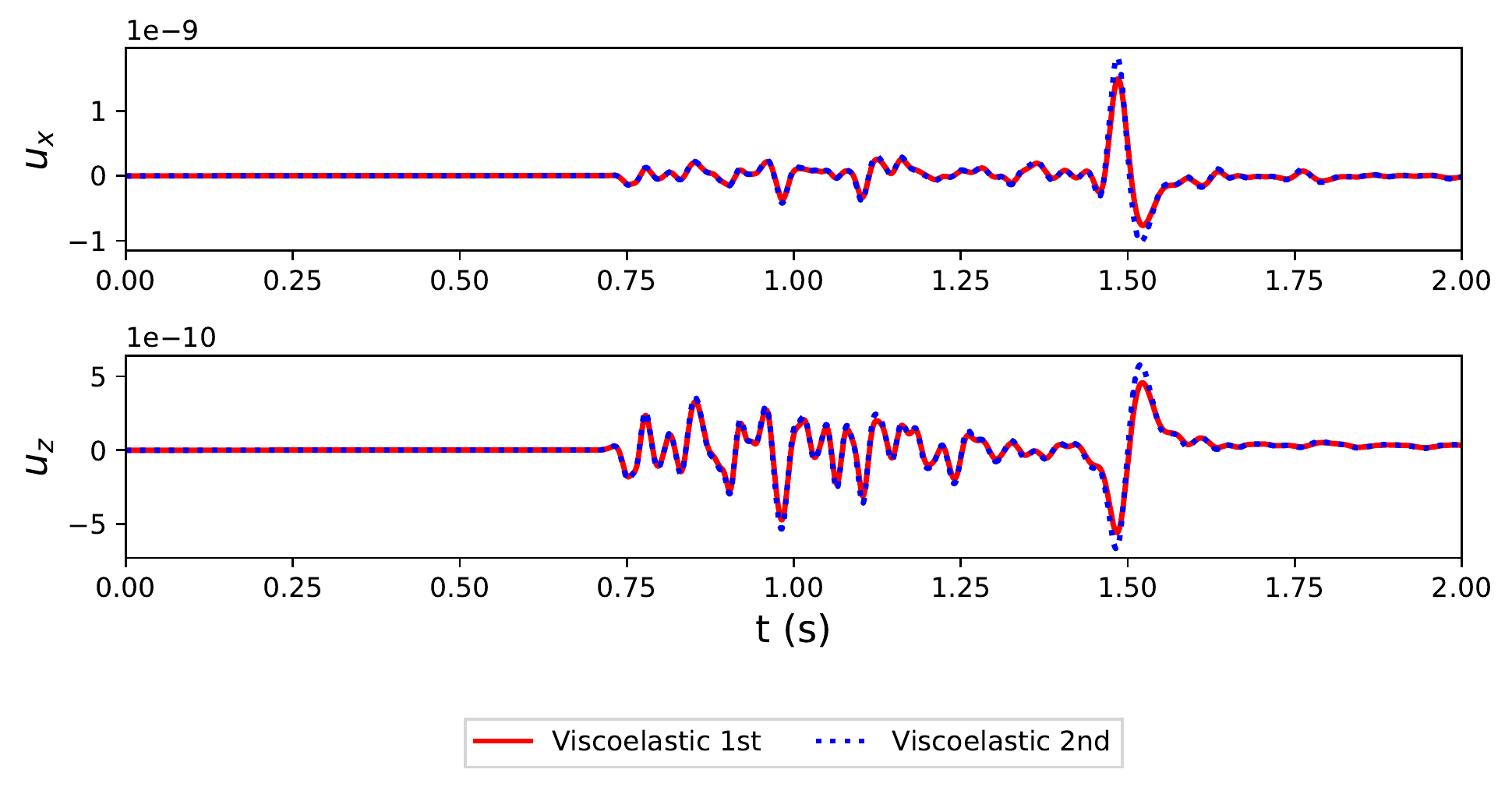}
\label{fig:left2}} 
\caption{
Single-trace seismograms for the receiver at $x_{R}=0.2$~km. The legend abbreviations ``Elastic'', ``Viscoelastic 1st'' and ``Viscoelastic 2nd'' correspond to the elastic wave equations ($Q_{P}=Q_{S}=\infty$) and the first- and second-order nearly constant $Q$ wave equations, respectively. Plots (a) and (b) show the same viscoelastic waveforms at different scales.     
}
\label{fig:single_left}
\end{figure}

\begin{figure}[hbt!]
\centering
\subfloat[Elastic and viscoelastic waves]
{\includegraphics[width=7.3cm]{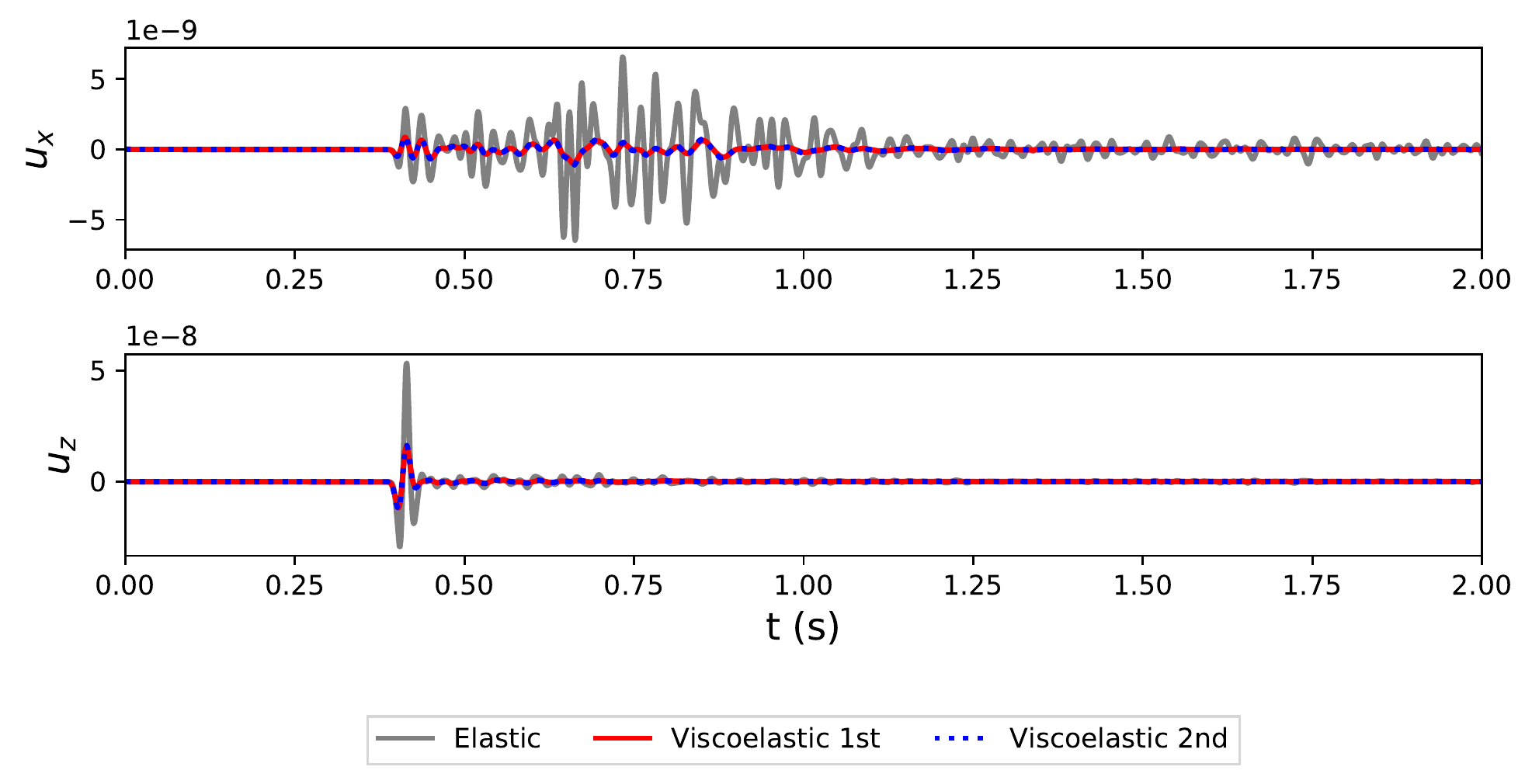}
\label{fig:center1}} 
\subfloat[Viscoelastic waves]
{\includegraphics[width=7.3cm]{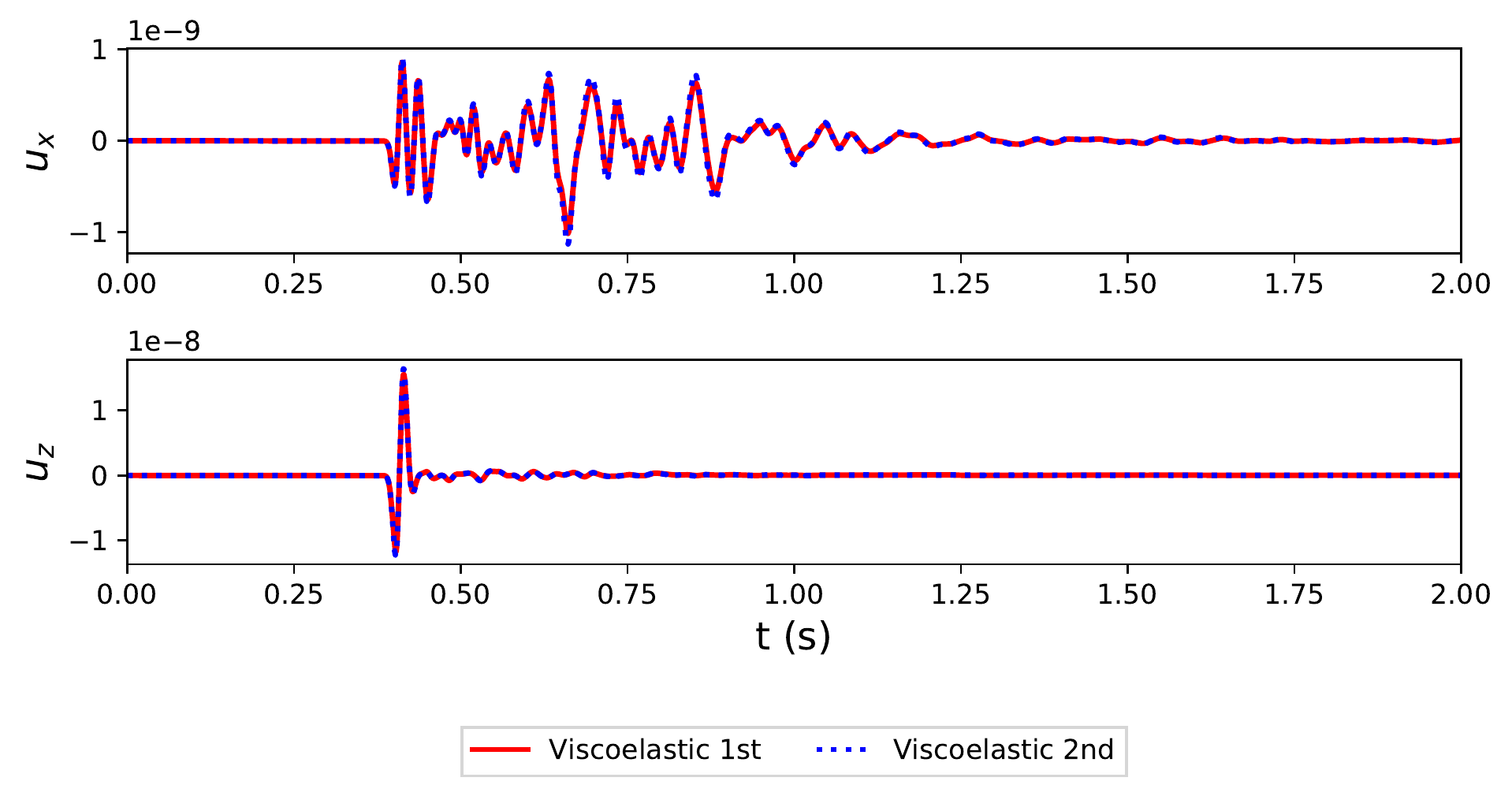}
\label{fig:center2}} 
\caption{
Similar to Figure \ref{fig:single_left} but for the receiver at $x_{R}=1.645$~km.    
}
\label{fig:single_center}
\end{figure}

\begin{figure}[hbt!]
\centering
\subfloat[Elastic and viscoelastic waves]
{\includegraphics[width=7.3cm]{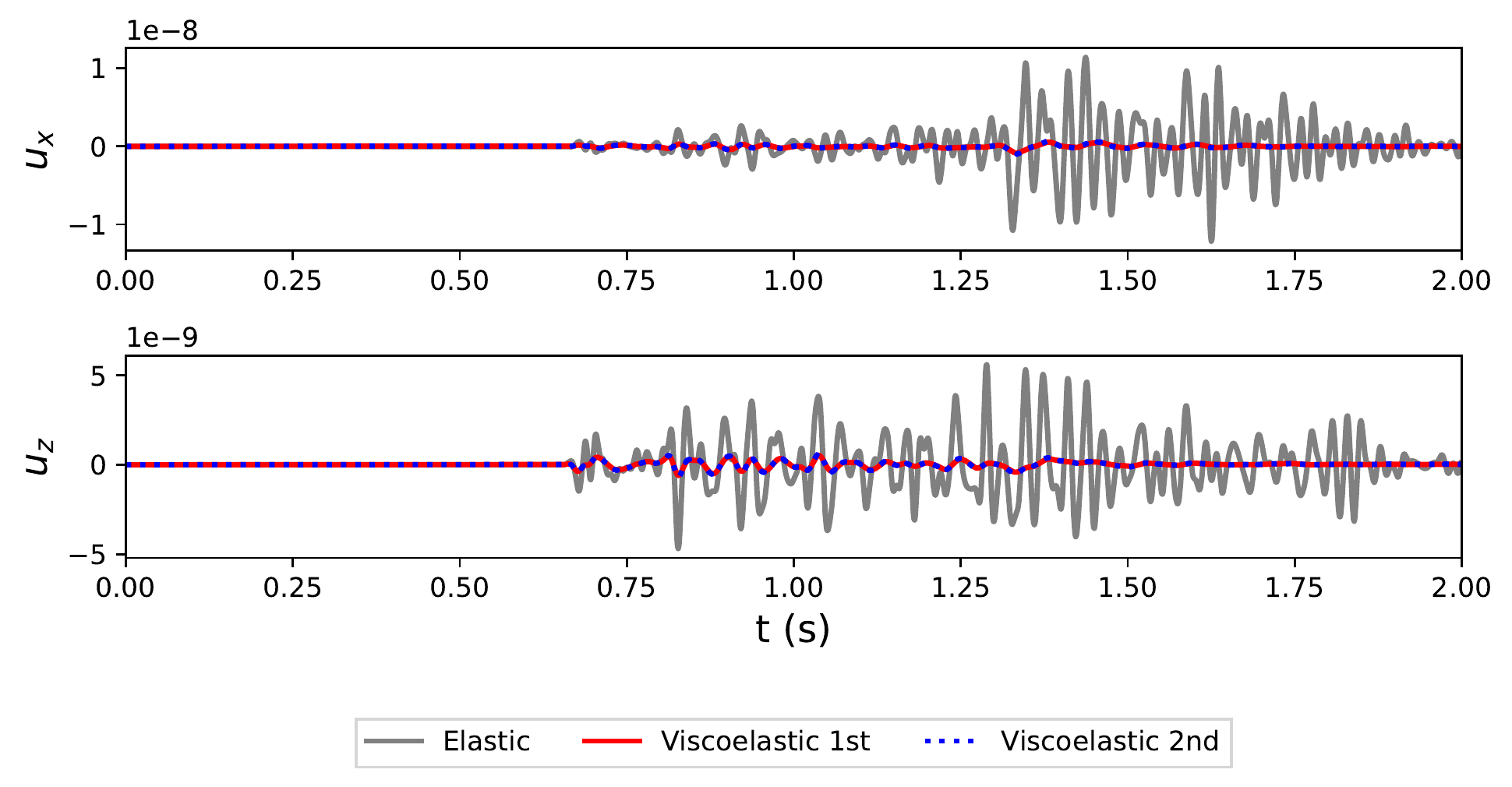}
\label{fig:right1}} 
\subfloat[Viscoelastic waves]
{\includegraphics[width=7.3cm]{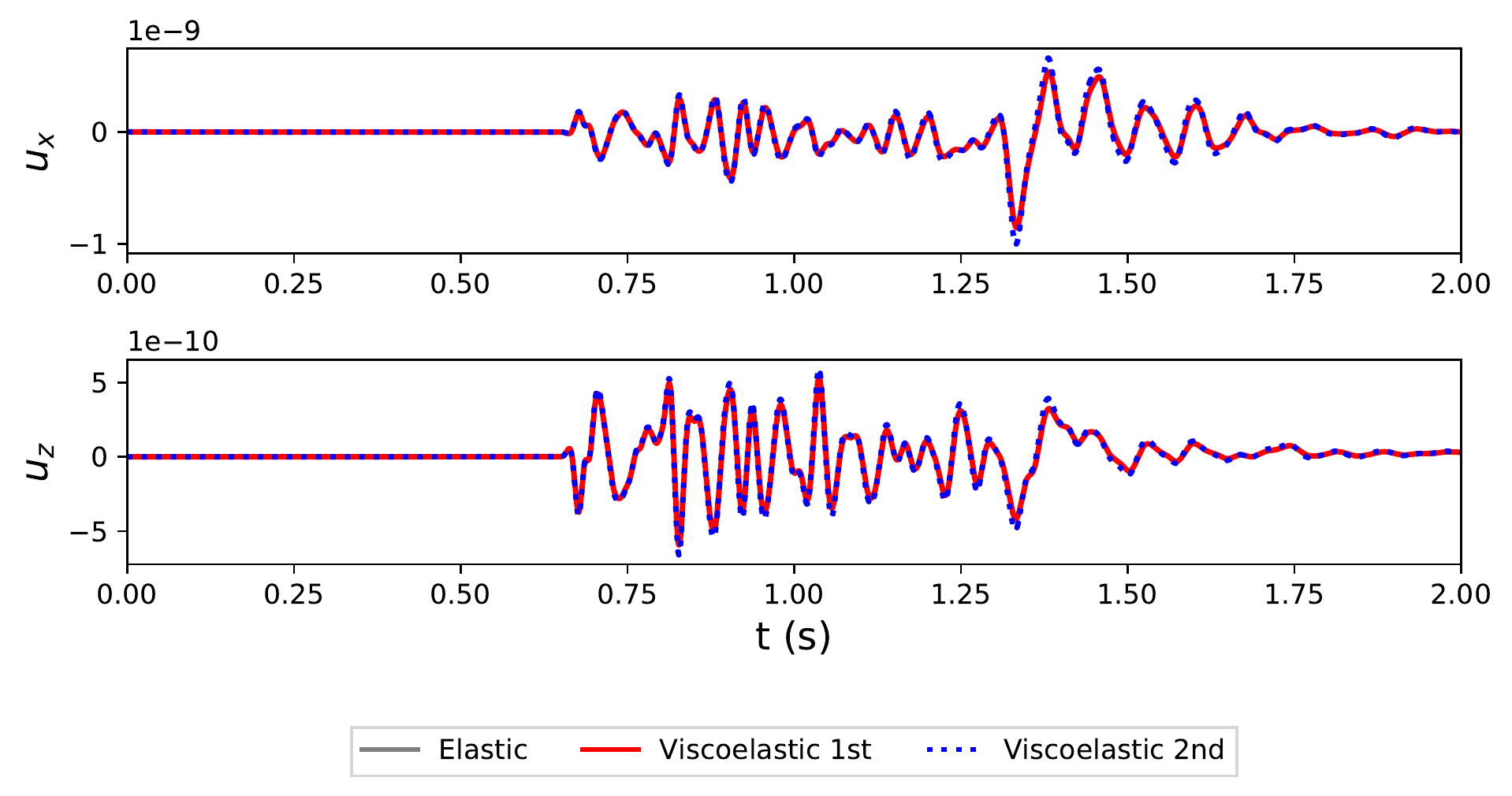}
\label{fig:right2}} 
\caption{
Similar to Figure \ref{fig:single_right} but for the receiver at $x_{R}=3.08$~km. 
}
\label{fig:single_right}
\end{figure}

We design a viscoelastic isotropic version of the Marmousi model. Using the P-wave model parameters in Figure \ref{fig:marm}, we set the S-wave model parameters as $v_{S}=v_{P}/2$ and $Q_{S}=7Q_{P}/10$. The reference frequency is set as $f_{0}=40$~Hz. The z-component of the source function $\mathbf{f}$ is initiated at $x=1.645$~km and $z=0.925$~km by a 40~Hz Ricker wavelet. According to the scaling property of the nearly constant $Q$ models \cite{hao.greenhalgh:2021}, the relaxation times in Table \ref{tab:tabl1} are divided by 0.65 so that the valid frequency range of the weighting function (equation \ref{eq:Weq}) is cast to $[0.65, 130]$~Hz, which can cover the frequency range of the Ricker wavelet. The receivers are located at the surface of the model. We use the finite-difference method \cite{emmerich:1987} to solve the wave equations in the 2-D $[x,z]$ plane. The Marmousi model is surrounded by a perfectly matched layer \cite{drossaert:2007} to absorb the artificial boundary reflections. Figures \ref{fig:single_left} through \ref{fig:single_right} indicate the decay of wave amplitude due to viscoelasticity. The waveform difference between the the first- and second-order nearly $Q$ models is small and visible, which is mainly caused by the quality factor difference between these two models, as we deduce from Figures \ref{fig:Qp} through \ref{fig:Vs2}). The details of this numerical example are shown in the supplementary material.  

\section{Conclusions}
The first- and second-order nearly constant $Q$ models are the approximations in a frequency range of interest to the Kolsky and Kjartansson models, respectively. The complex stiffness coefficients for the first- and second-order nearly constant $Q$ models share a $Q$-independent weighting function, the determination of which is dependent only on a frequency range of interest.  
The viscoelastic anisotropic wave equations for the first- and second-order nearly constant $Q$ models can be expressed in differential equation form and they explicitly involve the specified $Q$ parameters. The wave equations for these two models can be solved effectively by most of the existing time-domain wavefield modeling methods. 

\vskip6pt
\enlargethispage{20pt}

\section{Data and material availability}
Data, high-quality figures and plotting code can be accessed online at \url{http://github.com/xqihao/constQANI}.

\appendix
\section{The Kjartansson and Kolsky models}
In this Appendix, we show the complex stiffness coefficients in the Voigt notation for the Kjartansson and Kolsky models in the general viscoelastic anisotropic case. The Kjartansson and Kolsky models in the viscoacoustic case can be found in \cite{kjartansson:1979} and \cite{kolsky:1956}, respectively. 
According to \cite{hao.greenhalgh:2021}, the Kolsky model is the first-order approximation to the Kjartansson model. 

The non-zero independent elements of the complex stiffness coefficient matrix for the Kjartansson model are given by
\begin{equation} \label{eq:Mkjar}
M_{ij}(\omega) = M_{ij}^{(0)} \left(-i\frac{\omega}{\omega_{0}} \right)^{2\gamma_{ij}} ,
\end{equation}
with
\begin{equation} \label{eq:gamma_kjar}
\gamma_{ij} = \frac{1}{\pi}\arctan \left( \frac{1}{Q_{ij}} \right) ,
\end{equation}
where $\omega_{0}$ denotes a reference angular frequency. Quantity $\Gamma(.)$ denotes the Gamma function \cite{arfken:2013}. Quantities $M_{ij}^{(0)}$ denote reference stiffness coefficients corresponding to $Q_{ij} = \infty$. The minus sign in front of the imaginary unit ``$i$'' corresponds to the definition of the Fourier transform in equation \ref{eq:Fourier}. 

The Maclaurin series expansion of equation \ref{eq:Mkjar} with respect to $1/Q_{ij}$ is written as
\begin{equation} \label{eq:Mkjar_appr}
\frac{M_{ij}}{M_{ij}^{(0)}} = 1 + 
\frac{1}{Q_{ij}}
\left[\frac{2}{\pi} \text{ln} \left|\frac{\omega}{\omega_{0}}\right| - i \text{sgn}(\omega) \right] 
+ \frac{1}{2Q_{ij}^2} 
\left[\frac{2}{\pi} \text{ln}\left|\frac{\omega}{\omega_{0}}\right| - i \text{sgn}(\omega) \right]^2  +  
O\left(\frac{1}{Q_{ij}^3}\right) .
\end{equation}

Furthermore, the linear approximation of eq. \eqref{eq:Mkjar} with respect to $1/Q_{ij}$ leads to the non-zero independent elements of the complex stiffness coefficient matrix for the Kolsky model, namely
\begin{equation} \label{eq:Mkolsky}
M_{ij}(\omega) = M_{ij}^{(0)} \left \{
1 + \frac{1}{Q_{ij}}
\left[\frac{2}{\pi} \text{ln}\left|\frac{\omega}{\omega_{0}}\right| - i \text{sgn}(\omega) \right]  
\right \} .
\end{equation}
\section{The creep function matrices for the first- and second-order nearly constant $Q$ models}
This Appendix shows the derivation of the creep function matrices \ref{eq:X1stgen} and \ref{eq:X2ndgen}. 

As illustrated in equation \ref{eq:M1stgen}, the complex stiffness coefficient matrix for the first-order nearly constant $Q$ model is given by
\begin{equation}
\label{eq:M1stgen_app}
\mathbf{M}(\omega) = \mathbf{M}^{(0)} + \mathbf{M}^{(1)} \left[W(\omega) - W_{R}(\omega_{0}) \right] .
\end{equation}

The inverse of equation \ref{eq:M1stgen_app} gives rise to the complex compliance matrix for the first-order nearly constant $Q$ model, namely
\begin{equation} \label{eq:J1st}
\mathbf{J}(\omega) = \mathbf{J}^{(0)} \left\{\mathbf{I} + \mathbf{K}^{(1)} \left[W(\omega) - W_{R}(\omega_{0}) \right] \right\}^{-1} ,
\end{equation}
where matrices $\mathbf{J}^{(0)}$ and $\mathbf{K}^{(1)}$ are given by
\begin{align} 
\label{eq:J0new}
& \mathbf{J}^{(0)} = \left(\mathbf{M}^{(0)} \right)^{-1} , \\
\label{eq:K1new}
& \mathbf{K}^{(1)} = \mathbf{M}^{(1)} \mathbf{J}^{(0)} .
\end{align}

Referring to \cite{shalit:2017}, $\mathbf{J}$ can be expanded in a Neumann series
\begin{equation}
\mathbf{J}(\omega) = \mathbf{J}^{(0)}\sum_{n=0}^{\infty} (-1)^n \left(\mathbf{K}^{(1)}\right)^n 
\left[W(\omega) - W_{R}(\omega_{0}) \right]^n .
\end{equation}
According to \cite{horn:2012}, this Neumann series requires the condition given by
\begin{equation}
\parallel \mathbf{K}^{(1)} \left[W(\omega) - W_{R}(\omega_{0}) \right] \parallel_{1} < 1 ,
\end{equation}
where $\parallel \cdot \parallel_{1}$ denotes the 1-norm. 

If function $W(\omega) - W_{R}(\omega_{0})$ is viewed as a complex compliance, the corresponding creep function is given by $\zeta(t)$ in equation \ref{eq:zeta}. The creep function matrix for the first-order nearly constant $Q$ model is given by
\begin{equation}
\mathbf{X}(t) = \mathbf{J}^{(0)} \sum_{n=0}^{\infty} (-1)^n \left(\mathbf{K}^{(1)}\right)^n \zeta^{\langle n \rangle}(t) ,
\end{equation}
where $\zeta^{\langle n \rangle}(t)$ is defined as
\begin{equation} \label{eq:zetan}
\zeta^{\langle n \rangle}(t) =
\begin{cases}
\underbrace{\zeta(t) \odot \zeta(t) \cdots \odot \zeta(t)}_{n}, & \text{if } n > 1, \\
\zeta(t), & \text{if } n = 1 , \\
H(t), & \text{if } n = 0 .
\end{cases}
\end{equation}

As illustrated in equation \ref{eq:M2ndgen}, the complex stiffness coefficient matrix for the second-order nearly constant $Q$ model is given by
\begin{equation}
\label{eq:M2ndgen_app}
\mathbf{M}(\omega) = \mathbf{M}^{(0)} + \mathbf{M}^{(1)} \left[W(\omega) - W_{R}(\omega_{0}) \right] + \frac{1}{2} \mathbf{M}^{(2)} \left[W(\omega) - W_{R}(\omega_{0}) \right]^2 .
\end{equation}

The inverse of equation \ref{eq:M2ndgen_app} leads to the complex compliance matrix for the second-order nearly constant $Q$ model, namely
\begin{equation} \label{eq:J2ndnew}
\mathbf{J}(\omega) = \mathbf{J}^{(0)} \left\{ \mathbf{I} + \mathbf{K}^{(1)} \left[W(\omega) - W_{R}(\omega_{0}) \right] + \frac{1}{2} \mathbf{K}^{(2)} \left[W(\omega) - W_{R}(\omega_{0}) \right]^2 \right\}^{-1}  ,
\end{equation}
where $\mathbf{J}^{(0)}$ and $\mathbf{K}^{(1)}$ are given in equations \ref{eq:J0new} and \ref{eq:K1new}, respectively. Matrix $\mathbf{K}^{(2)}$ is given by
\begin{equation}  \label{eq:K2new}
\mathbf{K}^{(2)} =  \mathbf{M}^{(2)} \mathbf{J}^{(0)} .
\end{equation} 

The Neumann series of equation \ref{eq:J2ndnew} is given by
\begin{equation}
\mathbf{J}(\omega) = \mathbf{J}^{(0)} \sum_{n=0}^{\infty} (-1)^n
\left\{\mathbf{K}^{(1)} \left[W(\omega) - W_{R}(\omega_{0}) \right] + \frac{1}{2} \mathbf{K}^{(2)} \left[W(\omega) - W_{R}(\omega_{0}) \right]^2 \right\}^{n} ,
\end{equation}
where the validity condition is given by
\begin{equation}
\parallel
\mathbf{K}^{(1)} \left[W(\omega) - W_{R}(\omega_{0}) \right] + \frac{1}{2} \mathbf{K}^{(2)} \left[W(\omega) - W_{R}(\omega_{0}) \right]^2
\parallel_{1} < 1 .
\end{equation}

Utilizing the correspondence relation between function $W(\omega) - W_{R}(\omega_{0})$ and $\zeta(t)$, the creep function matrix for the second-order nearly constant $Q$ model is given by
\begin{equation}
\mathbf{X}(t) =  \mathbf{J}^{(0)} \sum_{n=0}^{\infty} (-1)^n
\left\{\mathbf{K}^{(1)} \zeta(t) 
+ \frac{1}{2} \mathbf{K}^{(2)} \zeta^{\langle 2 \rangle}(t) \right\}
^{\langle n \rangle} ,
\end{equation}
where superscript ``$\langle \cdot \rangle$'' is explained in equation \ref{eq:zetan}.

\vskip2pc
\bibliographystyle{./macros/elsarticle-num}
\bibliography{./refs/refs20210121,./refs/qi_refs20200812}

\end{document}